\pdfoutput=1
\documentclass[acmsmall,screen]{acmart}
\newif\ifextendedreport
\ifdefined\buildextendedreport \extendedreporttrue \fi

\AtBeginDocument{%
  }

\setcopyright{acmlicensed}
\copyrightyear{2018}
\acmYear{2018}
\acmDOI{XXXXXXX.XXXXXXX}

\acmConference[Conference acronym 'XX]{Make sure to enter the correct
  conference title from your rights confirmation email}{June 03--05,
  2018}{Woodstock, NY}
\acmISBN{978-1-4503-XXXX-X/2018/06}

\usepackage{longtable}
\usepackage{booktabs}
\usepackage{array}
\usepackage{hyperref}
\usepackage{tabularx}
\usepackage{multirow}
\usepackage[most]{tcolorbox}
\usepackage{booktabs}  
\usepackage{xcolor}    
\usepackage{colortbl}  
\usepackage{makecell}  
\usepackage{fontawesome5}
\usepackage{graphicx}  
\usepackage{acronym}
\usepackage{listings}
\usepackage{xcolor} 
\usepackage{booktabs}
\usepackage{tikz}
\usetikzlibrary{shadows,positioning,arrows.meta,calc,shapes,fit,backgrounds,decorations.pathreplacing}
\usepackage{graphics}
\usepackage{xspace}
\usepackage{comment}
\usepackage[frozencache]{minted} 
\usepackage{bchart}
\usepackage[htt]{hyphenat}
\usepackage{xspace}
\usepackage{pgfplots}
\usepackage{bm}
\usepackage{enumitem}
\usepackage{tabularx}
\usepackage{booktabs}

\usepackage[normalem]{ulem}

\usepackage[utf8]{inputenc}
\usepackage{textcomp}
\DeclareUnicodeCharacter{2265}{\ensuremath{\geq}}

\usepgfplotslibrary{statistics}
\pgfplotsset{compat=1.18}
\usepackage{adjustbox}
\usepackage{float}
\usepackage{wrapfig} 
\newcolumntype{R}[1]{>{\raggedleft\arraybackslash}p{#1}}
\newcolumntype{L}[1]{>{\raggedright\arraybackslash}p{#1}}

\acrodef{AI}{Artificial Intelligence} \acrodef{DL}{Deep learning}
\newcommand{\hbfa}{{\sc hbfa}\xspace}
\acrodef{AIA}{AI Accelerator} \acrodef{EAIA}{Edge AI Accelerator} \acrodef{EAIAs}{Edge AI Accelerators}
\acrodef{AP}{Application Processor}
\acrodef{UEFI}{Unified Extensible Firmware Interface}
\acrodef{SOC}{System On Chip} \acrodef{ISA}{Instruction set architecture}

\acrodef{MSRs}{Model Specific Registers} \acrodef{APIC}{Advanced Programmable Interrupt Controller}

\acrodef{MMIO}{Memory mapped I/O} \acrodef{IOMMU}{Input/Output Memory Management Unit}
\acrodef{MMU}{Memory Management Unit} \acrodef{AIAMMU}{\ac{AIA} Memory Management Unit}
\acrodef{PASID}{Process Address Space ID} \acrodef{IOTLB}{I/O Translation Lookaside Buffer}
\acrodef{SIMD}{Single Instruction Multiple Data} \acrodef{PCI}{Peripheral Component Interconnect}

\acrodef{OS}{Operating System} \acrodef{RM}{Reserved Memory} \acrodef{DTS}{Devicetree Source}
\acrodef{DTB}{Device Tree Blob} \acrodef{DTC}{Device Tree Compiler} \acrodef{TRM}{Technical Reference Manual}
\acrodef{KD}{Kernel Driver} \acrodef{KM}{Kernel Memory} \acrodef{HK}{Host Kernel}
\acrodef{NPU}{Neural Processing Unit} \acrodef{CDA}{Confused Deputy Attack}
\acrodef{UA}{Userspace Application} \acrodef{US}{User Space} \acrodef{USE}{User Space Entity}
\acrodef{SMP}{Shared Memory Page} \acrodef{SMID}{Shared Memory ID} \acrodef{USRT}{Userspace Space Runtime}
\acrodef{KS}{Kernel Space} \acrodef{TEE}{Trusted Execution Environment} \acrodef{LLM}{Large Language Model}
\acrodef{TPU}{Tensor Processing Unit} \acrodef{GPU}{Graphic Processing Unit}
\acrodef{GPGPU}{General Purpose \ac{GPU}} \acrodef{MAC}{Multiply Accumulate}
\acrodef{MAD}{Multiply Add} \acrodef{MMA}{Matrix Multiply Accelerator} \acrodef{TOPS}{Trillions Operations per Second}
\acrodef{FD}{File Descriptor}\acrodef{DLA}{Deep Learning Accelerator}

\acrodef{IAIA}{Inference AI Accelerator} \acrodef{ASP}{Application Specific Processor}
\acrodef{AVL}{AIA Vendor Library} \acrodefplural{AVL}{AIA Vendor Libraries} \acrodef{AF}{AI Framework}
\acrodef{AFRT}{AI Framework Runtime} \acrodef{AUA}{AI Userspace Application} \acrodef{ASIC}{Application Specific Integrated Circuit}

\acrodef{SMem}{System Memory} \acrodef{HMem}{Host Memory} \acrodef{DMem}{DMA Memory}
\acrodef{UMem}{User Memory} \acrodef{KMem}{Kernel Memory} \acrodef{AIMem}{AIA MMIO Memory}
\acrodef{AIRMem}{AIA Reserved Memory} \acrodef{DMA}{Direct Memory Access} \acrodef{CMA}{Contiguous Memory Allocation}
\acrodef{BAR}{Base Address Register} \acrodef{USB}{Universal Serial Bus}
\acrodef{IPC}{Inter Process Communication} \acrodef{POC}{Proof Of Concept}

\acrodef{LLMSK}{LLM Speed Kills Framework}
\acrodef{LLM}{Large Language Model}
\acrodef{PASID}{Process Address Space Identifier}
\acrodef{CST}{Concrete Syntax Tree}
\acrodef{UART}{Universal Asynchronous Receiver/Transmitter}

\definecolor{usermemcolor}{rgb}{1.0, 0.40, 0.40}
\definecolor{kernelmemcolor}{rgb}{0.40, 0.80, 0.0}

\acrodef{TFLite}{TensorFlow Lite} \acrodef{ORT}{ONNX Runtime}

\newcommand{\parahead}[1]{\noindent\textbf{#1}.}

\newcommand{\tbl}[1]{Table~\ref{#1}}
\newcommand{\sect}[1]{\S~\ref{#1}}

\newcommand{\fig}[1]{Figure~\ref{#1}}
\newcommand{\lst}[1]{Listing~\ref{#1}}

\newcommand{\apdxmissing}{\PackageWarning{macros}{Appendix pointer outside \string\extonly/\string\extelse}[appendix]}
\newcommand{\apdx}[1]{\ifextendedreport Appendix~\ref{#1}\else\apdxmissing\fi}
\newcommand{\apdxref}[1]{\ifextendedreport #1 in the appendix\else\apdxmissing\fi}
\newcommand{\Apdxref}[1]{\ifextendedreport #1 in the appendix\else\apdxmissing\fi}
\newcommand{\extonly}[1]{\ifextendedreport#1\fi}
\newcommand{\extelse}[2]{\ifextendedreport#1\else#2\fi}
\newcommand{\tobepublished}[1][Details]{\emph{[#1 withheld pending responsible disclosure; will be published soon.]}}

\newcommand{\eg}{\textit{e.g.,}\xspace}
\newcommand{\ie}{\textit{i.e.,}\xspace}
\newcommand{\etal}{\textit{et al.}\xspace}
\newcommand{\edk}{{\sc edk2}\xspace}
\newcommand{\afl}{{\sc afl++}\xspace}

\acrodef{UEFI}{Unified Extensible Firmware Interface}
\acrodef{ASLR}{Address Space Layout Randomization}
\acrodef{OS}{Operating System}
\acrodef{SMI}{System Management Interrupt}
\acrodef{SMM}{System Management Mode}
\acrodef{ASAN}{Address Sanitizer}
\acrodef{UBSan}{Undefined Behavior Sanitizer}

\newcommand{\systemname}{\textsc{Metis}\xspace}
\acrodef{ISA}{Instruction Set Architecture}
\acrodef{HBFA}{Host Based Firmware Analyzer}

\acrodef{CR}{Complete Rehosting}
\acrodef{MAR}{Minimal Abstraction Rehosting}
\acrodef{RAG}{Retrieval Augmented Generation}

\newcommand{\inputlib}{{\tt FuzzContextLib}\xspace}

\newcommand{\testap}{{\sc ea}\xspace}
\newcommand{\testapproto}[1]{\textsc{ea}\textsubscript{#1}\xspace}

\newcommand{\lemix}{\textsc{Lemix}\xspace}
\newcommand{\fuzzware}{\textsc{FuzzWare}\xspace}
\newcommand{\metislib}{{\tt MetisLib}\xspace}

\newcommand{\targetprotcol}{\ensuremath{P}\xspace}
\newcommand{\targetcomponent}{\ensuremath{C}\xspace}
\newcommand{\targetprotocol}{\ensuremath{P}\xspace}

\newcommand{\protocolfunctionset}[1]{\ensuremath{F_{#1}}\xspace}

\newcommand{\protocolfunction}[2]{\ensuremath{F^{#1}_{#2}}\xspace}

\newcommand{\dependenciesforp}[1]{\ensuremath{D(#1)}\xspace}

\newcommand{\dependenciesforpatlevel}[2]{\ensuremath{D^{#1}({#2})}\xspace}

\newcommand{\harnessforprotocol}[1]{\ensuremath{H_{#1}}\xspace}
\newcommand{\mockforprotocol}[1]{\ensuremath{M(#1)}\xspace}
\newcommand{\highfidelity}{{\ensuremath{HF}}\xspace}
\newcommand{\lowfidelity}{{\ensuremath{LF}}\xspace}
\newcommand{\highfidelitymock}[1]{{\ensuremath{HF(#1)}}\xspace}
\newcommand{\lowfidelitymock}[1]{{\ensuremath{LF(#1)}}\xspace}

\newcommand{\interfacedepsmap}[1]{\ensuremath{\psi_{#1}: \protocolfunctionset{#1} \rightarrow (\protocolfunctionset{#1} \mid \emptyset{})}\xspace}

\newcommand{\interfacedeps}[1]{\ensuremath{\psi_{#1}}\xspace}

\newcommand{\interfacedepsfunc}[2]{\ensuremath{\psi_{#1}(#2)}\xspace}

\acrodef{DCR}{Direct-dependency Component Rehosting}

\newcommand{\numprotocols}{162\xspace}
\newcommand{\numdrivers}{60\xspace} 

\newcommand{\add}{\textcolor{red}{XXX}\xspace}
\newcommand{\numpreviousbugs}{14\xspace} 
\newcommand{\avgcodecvgperc}{73.1\%\xspace} 
\newcommand{\percfnrate}{71.4\%\xspace} 

\newcommand{\remotebugs}{12\xspace} 
\newcommand{\numreportedbugs}{26\xspace} 

\newcommand{\avgfncvgperc}{85.5\%\xspace} 
\newcommand{\numrediscoveredbugs}{4\xspace} 
\newcommand{\numafluniquecrashes}{5,116\xspace} 
\newcommand{\numdistinctcrashes}{204\xspace} 
\newcommand{\numdriverbugs}{137\xspace} 
\newcommand{\cvgimprovement}{1.7$\times$\xspace} 
\newcommand{\numfuzzuerprotocols}{55\xspace} 
\newcommand{\numfuzzuerbugs}{8\xspace} 
\newcommand{\hbfalinecvgperc}{15.2\%\xspace} 
\newcommand{\metispartitionlinecvgperc}{62.6\%\xspace} 
\newcommand{\llmfncvgperc}{85.5\%\xspace} 
\newcommand{\llmlinecvgperc}{79.2\%\xspace} 
\newcommand{\numllmworking}{25\xspace} 
\newcommand{\numllmmeasured}{25\xspace} 
\newcommand{\numllmbugs}{32\xspace} 
\newcommand{\metislinecvgonllmset}{76.7\%\xspace} 
\newcommand{\metisfncvgonllmset}{89.1\%\xspace} 

\newcommand{\fuzzuer}{{\sc FuzzUEr}\xspace}

\tcbset{
  observation/.style={
    colback=yellow!20!white,
    colframe=black,
    boxrule=0.4pt,
    left=2pt,right=2pt,top=2pt,bottom=2pt,
    arc=2pt,
    fonttitle=\bfseries,
    before skip=4pt, after skip=4pt,
  }
}

\tcbset{
  openproblem/.style={
    colback=blue!20!white,
    colframe=black,
    boxrule=0.4pt,
    left=2pt,right=2pt,top=2pt,bottom=2pt,
    arc=2pt,
    fonttitle=\bfseries,
    before skip=4pt, after skip=4pt,
  }
}

\newif\ifshowcomments
\showcommentstrue
\ifshowcomments
\newcommand{\connor}[1]{\textcolor{blue}{(Connor: #1)}}
\newcommand{\machiry}[1]{\textcolor{green}{\textbf{Machiry:} #1}}

\newcommand{\srihari}[1]{\textcolor{purple}{\textbf{Srihari} #1}}
\newcommand{\aditya}[1]{\textcolor{CadetBlue}{\textbf{Aditya} #1}}

\else
\newcommand{\connor}[1]{\ignorespaces}
\newcommand{\machiry}[1]{\ignorespaces}

\newcommand{\srihari}[1]{\ignorespaces}
\newcommand{\aditya}[1]{\ignorespaces}

\fi
\newcommand{\strike}[1]{\sout{#1}}

\definecolor{forestgreen}{RGB}{34, 139, 34}

\definecolor{limitcolor}{rgb}{1.0, 0.49, 0.0}

\definecolor{nocontrolcolor}{rgb}{0.54, 0.25, 0.27}

\newcommand{\inlinecode}[1]{%
\mintinline[fontsize=\small{},mathescape, escapeinside=||]{c}{#1}%
}

\newcommand{\whatweneed}[2]{
\begin{tcolorbox} [width=1.0\linewidth, colback=red!07!white, top=1pt, bottom=1pt, left=2pt, right=2pt]
\textbf{Challenge #1}: #2
\end{tcolorbox}
}

\newtcbtheorem[]{promptbox}{Prompt}
{
  enhanced,
  colback=blue!5!white,
  colframe=blue!40!white,
  coltitle=black,
  fonttitle=\bfseries\small,
  fontupper=\footnotesize,
  boxrule=0.2mm,
  arc=1mm,
  left=1mm,
  right=1mm,
  top=0.5mm,
  bottom=0.5mm,
  boxsep=0.5mm,
  before skip=4pt,
  after skip=4pt,
  separator sign={:},
  description font=\bfseries\small,
}{pr}
\begin{document}

\title{Towards Effective Fuzzing of EDK2 Components}

\author{Datta Manikanta Sri Hari Danduri}
\email{ddanduri@purdue.edu}
\author{Connor Everett Glosner}
\email{cglosne@purdue.edu}
\author{Aravind Machiry}
\email{amachiry@purdue.edu}
\affiliation{%
  \institution{Purdue University}
  \city{West Lafayette}
  \state{Indiana}
  \country{USA}
}


\begin{abstract}
Securing bootloaders is critical to establishing the trusted computing base of modern systems.
TianoCore's edk2, the de facto \ac{UEFI}-based bootloader, is widely deployed but has recently been shown to contain numerous high-severity vulnerabilities.
Existing approaches either focus on post-boot interfaces or rely on heavyweight simulation, limiting their scalability, throughput, and coverage of boot-time components.

We present Metis, an automated technique for fuzzing edk2 protocols (\ie{} boot-time interfaces) via rehosting as a user-space applications.
Our key insight is that effective fuzzing must exercise not only protocol interfaces but also their dependency interactions, which serve as additional input channels.
Metis employs a \ac{DCR}-based strategy to construct semantically meaningful mocks for dependencies, enabling controlled input injection across all channels.
It further leverages a hybrid approach combining static analysis and \ac{LLM}-guided synthesis to generate and refine fuzzing harnesses.
We implement Metis on \ac{HBFA} framework and evaluate it on \numdrivers{} drivers covering \numprotocols{} protocols.
Metis achieves higher coverage (\avgcodecvgperc{}, 
$\sim$\cvgimprovement{} improvement over \fuzzuer{} the state-of-the-art \edk{} fuzzing technique), 
and discovers \numdriverbugs{} previously unknown bugs.
We have already reported \numreportedbugs{} of them, including \remotebugs{} that are remotely triggerable, and all reported vulnerabilities have been acknowledged by the vendors.
\end{abstract}
\maketitle

\section{Introduction}
\label{sec:introduction}
Bootloaders \cite{matrosov2019rootkits} play a critical role in establishing the security of modern computing platforms, as they are responsible for initializing hardware and loading the operating system.
The \acf{UEFI} specification \cite{uefispec} has emerged as the dominant standard for bootloaders, and most commodity platforms implement \ac{UEFI}-compliant firmware.
\edk{} \cite{edk2sourcerepo} serves as the reference implementation of \ac{UEFI} and is widely adopted \cite{nvidiafork, armfork, microsoftfork, delldisclosure, lenovodisclosure, appledisclosure} by major vendors such as ARM, AMD, Intel, NVIDIA, Lenovo, Apple, Dell, Microsoft, etc.

Vulnerabilities in \edk{} can have severe consequences, enabling bootkits \cite{blacklotus, blacklotus_2} and other persistent attacks \cite{matrosov2019rootkits} that remain resilient even after reinstallation of the \ac{OS}.
Given its widespread deployment and critical role in establishing a trusted computing base, securing \edk{} is of paramount importance.
However, the current state of \edk{} is concerning.
Recent exploits \cite{binarlylogofail, PixieFail, blacklotus, blacklotus_2} have demonstrated the prevalence of severe vulnerabilities across various \edk{} components, commonly referred to as protocols.
The absence of widely adopted exploit mitigation mechanisms, such as \ac{ASLR} \cite{shacham2004effectiveness}, further exacerbates the problem, making these vulnerabilities easier to exploit in practice.

Dynamic analysis, particularly fuzz testing \cite{manes2019art}, has proven effective in uncovering vulnerabilities in complex software systems.
Prior fuzzing efforts \cite{simics-fuzzer, Yin2023rsfuzzer, mahesh2023validation, glosner_fuzzuer_2025, uefuzzer, baz2015symbolic, excite} targeting \edk{} have largely focused on post-boot interfaces, especially \ac{SMI} \cite{mouzakitis2021x86} handlers, due to their well-defined invocation mechanisms.
%
Existing efforts to test arbitrary \edk{} protocols remain limited in scope, often focusing on identifying and analyzing isolated components.
While such approaches are effective for relatively self-contained libraries, extending them to arbitrary \edk{} components with intricate dependencies requires substantial manual effort and is often impractical.

To address this, prior works such as Yang \etal{} \cite{simics-fuzzer} and \fuzzuer{} \cite{glosner_fuzzuer_2025} attempt a whole-system approach by executing the entire firmware image within a simulator and triggering individual components.
However, these approaches rely heavily on the availability of high-quality fuzzing harnesses.
Furthermore, the lack of sanitizers (\eg{}\ac{ASAN} \cite{serebryany2012addresssanitizer}, \ac{UBSan} \cite{ubsan, song2019sok}) in the boot environment further hinders bug detection.
The recent work, \fuzzuer{}, addressed these issues by designing an automated harness generation technique and implementing partial \ac{ASAN} support.
However, these harnesses fail to comprehensively capture the diverse input mechanisms of \edk{} protocols.
Consequently, as we demonstrate in \sect{sec:evaluation}, they are ineffective at thoroughly exercising and testing them.
Finally, adding support for additional sanitizers, such as \ac{UBSan}, requires significant engineering effort.

In this paper, we present \systemname{}, an automated fuzzing framework for arbitrary \edk{} protocols.
Inspired by recent techniques \cite{tanksalkar2025lemix, tay2023greenhouse, scharnowski2022fuzzware}, our approach rehosts \edk{} protocols as user-space applications, enabling the use of existing fuzzing tools (\eg{} \afl{} \cite{fioraldi2020afl}) and sanitizers without requiring any modifications.
However, existing rehosting approaches are ineffective when applied to \edk{} protocols.
Our key observation is that \edk{} protocols can be exercised through two primary interaction mechanisms: (i) their interface functions and (ii) their dependencies on other protocols and libraries.
To address this, we design a \emph{\ac{DCR}} technique that rehosts arbitrary \edk{} protocols by constructing appropriate mocks for their direct dependencies.
These mocks not only emulate the required execution environment but also provide controlled input channels through dependency interactions.
In addition, we generate harnesses that systematically translate fuzzer inputs into well-formed invocations of protocol interfaces.
\systemname{} operates in two phases: \emph{Rehosting} and \emph{Fuzzing}.
In the \emph{rehosting phase}, \systemname{} leverages a combination of static analysis and \ac{LLM}-based synthesis to automatically generate harnesses and mocks. 
It then composes the target protocol, generated harnesses and mocks, and a runtime library that emulates essential \edk{} services using standard Linux abstractions into a standalone user-space executable, referred to as \edk{} test application or \testap{}.
We also use a dynamic refinement technique to improve the harnesses and mocks using a small testing loop.
In the \emph{fuzzing phase}, \systemname{} executes \testap{} using, \afl{}, the state-of-the-practice coverage-guided fuzzer.
It further incorporates automated crash deduplication and filtering to eliminate false positives (\eg{} those arising from imprecise mocks), and reports validated bugs.

We evaluate \systemname{} on \numprotocols{} \edk{} protocols containing \numpreviousbugs{} previously known bugs.
Our results show that \systemname{} achieves an average of \avgcodecvgperc{} code coverage, providing a \cvgimprovement{} improvement over \fuzzuer{}.
Moreover, \systemname{} discovers \numdriverbugs{} previously unknown bugs, which we are in the process of reporting, as creating a proof-of-concept (PoC) for each takes time.
We have already reported \numreportedbugs{} of them, out of which \remotebugs{} are remotely triggerable, and all our reported vulnerabilities have been acknowledged by the vendors.

In summary, the following are our contributions:
\begin{itemize}[noitemsep, leftmargin=*]
  \item We design \systemname{}, that uses \ac{DCR} technique and enables fuzz testing of arbitrary \edk{} protocols by constructing user-space mocks for their direct dependencies.

  \item We developed techniques that combine static analysis with \ac{LLM}-based synthesis to automatically generate test harnesses and dependency mocks. We further refine these artifacts using feedback from dynamic analysis.

  \item Our extensive evaluation on \numprotocols{} \edk{} protocols containing \numpreviousbugs{} previously known vulnerabilities show that \systemname{} significantly outperforms existing approaches, achieving an average of \avgcodecvgperc{} code coverage.

  \item We discover \numdriverbugs{} previously unknown bugs and have already reported \numreportedbugs{} of them, including \remotebugs{} that are remotely triggerable; all reported vulnerabilities have been acknowledged by the vendors.
\end{itemize}

\section{Background and Notations}
\label{sec:background}

\extonly{This section introduces the background required to understand our technique.}


\subsection{\acf{UEFI}}
\label{subsec:uefi}

The \ac{UEFI} specification~\cite{uefi-spec} defines a standardized software interface between platform firmware and the operating system.
\ac{UEFI} firmware executes in several phases: Security (SEC), Pre-EFI Initialization (PEI), Driver Execution Environment (DXE), Boot Device Selection (BDS), and Runtime\extonly{, each with progressively richer execution capabilities}~\cite{beyond-bios}.\extonly{
The DXE phase is the most complex stage of the boot process and provides the core interfaces used by the \ac{OS} loader and boot device selection components to interact with the bootloader.
Additionally, the DXE phase is when the post-boot interfaces are created.

\subsubsection{Services}}
\ac{UEFI} exposes standardized interfaces through \emph{services} (\ie{} Boot Services and Runtime Services) and \emph{protocols}.\extonly{ 
These collectively define the functionality that any \ac{UEFI}-compliant bootloader must implement.
All \ac{UEFI} interfaces are accessed via the \inlinecode{EFI_SYSTEM_TABLE} pointer, which is passed as an argument to each component. 
Boot and Runtime Services are accessed through the \inlinecode{BootServices} and \inlinecode{RuntimeServices} fields of this table (\eg{} the \inlinecode{AllocatePages} boot service can be invoked via \inlinecode{ST->BootServices->}\inlinecode{AllocatePages} function pointer). 
The bootloader initializes these function pointers to their corresponding implementations.
\ac{UEFI} also specifies event callbacks, which enable asynchronous interactions.
For instance, a component can register timer events that are invoked periodically.

\subsubsection{Protocols}
While services provide a relatively fixed and standardized interface, \emph{protocols play a central role in decoupling implementation from interface}.} 
A protocol in \ac{UEFI} is a named interface, defined as a C structure (\emph{interface structure}) of function pointers (referred to as \emph{interface functions}) and identified by a 128-bit GUID.\extonly{
Unlike services, which are initialized during bootloader startup, protocols enable flexible, asynchronous interactions between components.} 
Any component (\eg{} a driver) can install a protocol using \texttt{InstallProtocolInterface} or \texttt{InstallMultipleProtocolInterfaces} boot services.\extonly{ 
\Apdxref{\lst{lst:dhcprotocolinstallation}} shows an example of installing a protocol, where \inlinecode{gEfiDhcp4ProtocolGuid} is the GUID and \inlinecode{mDhcp4ProtocolTemplate} is the structure containing function pointers to various functions.}
Similarly, components can discover and use protocols via services such as \texttt{LocateProtocol}, \texttt{OpenProtocol}, and \texttt{HandleProtocol}.
For example, the DHCP driver opens the UDP protocol with \texttt{OpenProtocol}.

\subsection{\edk{}}
\label{subsec:edk2}
TianoCore \edk~\cite{edk2sourcerepo} is the reference open-source implementation of \ac{UEFI} firmware.\extonly{
As discussed in \sect{sec:introduction}, \edk{} is the most widely deployed bootloader in commodity systems.}
It comprises over 1.2 million lines of C code~\cite{edk2checkedc} organized into packages\extonly{ (\eg{} \texttt{MdeModulePkg}, \texttt{NetworkPkg}, \texttt{SecurityPkg})}, each containing multiple components, \ie{} drivers and library classes.\extonly{
Each component (\ie{} driver or library) may implement one or more protocols, which are dynamically installed using a GUID and a set of function pointers for the interface structure.} 
The implementation of each protocol may depend on other protocols and libraries, and interact with them through \ac{UEFI} event callbacks.

\subsubsection{Protocol Interactions}
\label{subsec:interaction}

\edk{} protocols interact with other protocols and \ac{UEFI} entities primarily through the following mechanisms.\extonly{
We illustrate different mechanisms through our running example, the DHCP driver (\texttt{Dhcp4Dxe}).}

\begin{itemize}[leftmargin=*, noitemsep]

\item \textbf{\ac{UEFI} Services.}
Protocols invoke \ac{UEFI} services\extonly{ via the system table (\sect{subsec:uefi}), enabling access to core firmware functionality such as memory management, boot services, and runtime services}.
For example, the DHCP driver calls \inlinecode{gBS->RaiseTPL} and \inlinecode{gBS->OpenProtocol}, where \inlinecode{gBS} is the global variable pointing to the boot services.

\item \textbf{Protocol Dependencies.}
Protocols may depend on other protocols, either directly or transitively, to implement their functionality.
For instance, the DHCP protocol implementation depends on the UDP protocol ({\tt gEfiUdp4ProtocolGuid}).

\item \textbf{\ac{UEFI} Events.}
\extonly{As discussed in \sect{subsec:uefi}, }\ac{UEFI} supports asynchronous interactions through event callbacks.
Protocols may create or register events, enabling deferred or periodic execution.
For instance, the DHCP driver registers an event callback through the function {\tt UdpIoRecvDatagram}.

\item \textbf{Libraries.}
\edk{} introduces modularity through \emph{library classes}, which are abstract interfaces whose concrete implementations are selected at build time via the platform's DSC (Description) file.\extonly{
A module declares dependencies on library classes (e.g., \texttt{MemoryAllocationLib}), and the build system resolves them to platform-specific implementations (\eg{} x86 or ARM).}
Protocols may rely on such libraries to implement core functionality.\extonly{
For instance, the DHCP driver's call to the \inlinecode{AllocatePool} library function will be resolved to a platform-specific implementation of memory management functions.}

\end{itemize}

\subsubsection{Notations and Definitions}
\label{subsubsec:notationsanddefinitions}
We use the following notations and definitions to describe the key entities in our approach:
\begin{itemize}[leftmargin=*, noitemsep]
\item \textbf{Component and Protocols.} We use the term \emph{component} to denote a self-contained software module, such as a driver or library.
Each component may implement and install one or more protocols.
We use \targetprotcol{} to denote the target protocol under test, and \targetcomponent{} to denote the component that implements this protocol.
\item \textbf{Interface Functions:}
The set of all interface functions of \targetprotcol{} is represented by \protocolfunctionset{\targetprotcol{}}.
Individual interface functions are denoted using superscripts; \eg{} \protocolfunction{1}{\targetprotcol{}} and \protocolfunction{2}{\targetprotcol{}} represent two distinct interface functions of \targetprotcol{}.
For instance, \extonly{in \apdxref{\lst{lst:dhcprotocolinstallation}}, }\protocolfunction{1}{DHCP} = {\tt  EfiDhcp4TransmitReceive}, as it is an interface function for the protocol.

\item \textbf{Interface Dependencies:}
Interface functions within a protocol may exhibit ordering dependencies.
In particular, certain functions must be invoked before others to ensure correct execution; otherwise, functions may exit early.
For example, the DHCP protocol requires the \inlinecode{Configure} function to be called before any other interface function.
We use \interfacedepsmap{\targetprotcol{}} to denote the dependencies between interface functions, \ie{} 
\interfacedepsfunc{\targetprotcol{}}{f} is the set of interface functions in \targetprotocol{} (or empty set) which should be called before $f$.
\item \textbf{Protocol and Component Dependencies:}
We use \dependenciesforp{x} to denote the set of all dependencies of protocol $x$.
Dependencies at a specific level are denoted using superscripts.
For instance, \dependenciesforpatlevel{1}{x} represents the set of direct (Level~1) dependencies of $x$, while \dependenciesforpatlevel{2}{x} represents its Level~2 dependencies.
All dependencies of a target component are inherited by the protocols it implements.\extonly{
For example, \apdxref{\fig{fig:working-example}} shows the partial dependency tree for the {\tt Dhcp4Dxe} driver, which is shared by all protocols provided by this component, including {\tt DHCP4\_PROTOCOL} and {\tt DHCP4\_SERVICE\_BINDING\_PROTOCOL}.}

\item \textbf{Harness and Mocks:}
We use \harnessforprotocol{x} to denote the fuzzing harness for protocol $x$.
The harness consumes input data (\eg{} from standard input) and invokes the interface functions of $x$ accordingly.
Similarly, we use \mockforprotocol{x} to denote the mock for protocol $x$, \ie{} a semantically equivalent implementation of $x$ which uses input data to provide the output.

\end{itemize}

\extonly{\subsubsection{\ac{UEFI} Protocols v/s Network Protocols}
We want to emphasize that \ac{UEFI} protocols and traditional network protocols, despite sharing the term ``protocol,'' are fundamentally different.
\ac{UEFI} protocols are in-memory firmware interfaces, implemented as C structures of function pointers and identified by GUIDs, and are used for intra-firmware communication.
In contrast, network protocols (\eg{} DHCP, TCP/IP) define on-the-wire message formats and communication semantics between distributed systems.
Throughout this paper, the term \emph{protocol} refers to a \ac{UEFI} protocol.
In particular, our running example (\lst{lst:newrunningexample}) of DHCP refers to the \ac{UEFI} DHCP protocol, which provides DHCP functionality as a firmware interface for other components, and should not be confused with the network DHCP protocol.}

\subsection{Rehosting}
\label{subsec:rehosting}
\emph{Rehosting} \cite{10.1145/3433210.3453093} a bootloader, or more generally, firmware, refers to executing it in a non-native environment, typically within a virtualized or user-space setting.\extonly{
This approach enables scalable fuzzing by removing the dependence on specialized hardware and allowing fine-grained instrumentation and visibility into firmware execution.}

\subsubsection{\acf{HBFA}}
\label{subsec:hbfa}
\ac{HBFA}~\cite{hbfa} is an open-source framework for fuzzing \ac{UEFI} components in user space by rehosting \edk{} modules using host-side toolchains and stub libraries (\inlinecode{HostLibs}) that map \ac{UEFI} services to POSIX equivalents.\extonly{
This enables execution under coverage-guided fuzzers such as \afl{}~\cite{fioraldi2020afl} and libFuzzer~\cite{libfuzzer}.}
However, \emph{\ac{HBFA} provides limited support for \edk{} libraries: many stubs are empty or fail to preserve semantics.}\extonly{
For example, stateful behavior between \inlinecode{SetVariable} and \inlinecode{GetVariable} is not maintained, leading to inconsistent execution.}
Additionally, all protocol dependencies must be resolved at link time, requiring developers to manually implement mocks for each dependency.

As a result, \ac{HBFA} is primarily used to reproduce known vulnerabilities at the function level, with mocks created reactively for known CVEs rather than enabling systematic discovery of new bugs~\cite{glosner_fuzzuer_2025}.

\section{Motivation}
\label{sec:motivation}


Previous work \cite{glosner_fuzzuer_2025} shows that over 70\% of previously known vulnerabilities in \edk{} reside in protocol implementations.
Motivated by this observation, we aim to \emph{enable effective fuzz testing of \edk{} protocols}.
Scalable fuzzing of \edk{} protocols requires two key capabilities:
(1) the ability to execute protocols in a hardware-independent environment, and
(2) the ability to supply inputs through all of their interaction channels.
While existing techniques address the first requirement through simulation-based execution, they fall short in satisfying the second.
In particular, they are unable to provide inputs through protocol dependencies and library interactions, which constitute a significant portion of the attack surface.
For example, \fuzzuer{} adopts a simulation-based approach that enables hardware-independent execution but exercises the target protocol only through its interface functions.
As a result, it cannot effectively fuzz protocols such as DHCP, which consume inputs indirectly through other protocols (\eg{} UDP).
Since \fuzzuer{} interacts with such protocols solely via their interfaces, it fails to deliver inputs through these dependency channels.
As we show in \sect{sec:evaluation}, this limitation fundamentally restricts \fuzzuer{}'s code coverage and hinders comprehensive vulnerability discovery.

\section{\systemname{}}

\begin{wrapfigure}[9]{R}{0.5\textwidth}
\vspace{-\baselineskip}
\fbox{\parbox[c][4\baselineskip][c]{\dimexpr\linewidth-2\fboxsep-2\fboxrule\relax}{\centering\small\tobepublished[Code]}}
\vspace{-0.5\baselineskip}
\captionof{listing}{\footnotesize Running example: the DHCP driver (\texttt{Dhcp4Dxe}) of \edk{}, in which \systemname{} discovered two real vulnerabilities.}
\label{lst:newrunningexample}
\vspace{-\baselineskip}
\end{wrapfigure}

Existing rehosting approaches fail to comprehensively exercise \edk{} protocols.
We observe that \emph{a protocol (\targetprotocol{}) primarily interacts through its interface functions and its direct dependencies}, \eg{} DHCP primarily depends on UDP and RNG, \ie{} its \dependenciesforpatlevel{1}{\targetprotocol{}}.
Thus, in principle, \emph{exercising through direct dependencies and interface functions is sufficient to achieve comprehensive coverage}.
However, dependent protocols themselves introduce additional dependencies, creating deep interaction chains. \eg{} UDP (direct dependency of DHCP) depends on IP.

A natural solution is to create models for the direct dependencies, which preserve the interface, \eg{} through mock objects \cite{thomas2002mock}.
Yet, unless these mocks faithfully preserve semantics, they can introduce infeasible behaviors \cite{muralee2025reactivebottomuptesting}, leading to false positive crashes that are not realizable in real \edk{} executions. We also need a capability to filter out such spurious crashes.


Based on this insight, we propose \acf{DCR}, a technique that rehosts a protocol by constructing mocks for its direct dependencies.
By focusing only on direct dependencies, \ac{DCR} avoids the constraints imposed by deeper dependency chains.
Next, we identify external interaction points (e.g., protocol calls, library invocations, and timer callbacks) and replace them with stubs driven by fuzzer input, enabling control over all protocol input channels.
Finally, we generate harnesses for protocol interface functions and combine them with the mocks and target protocol into a user-space \edk{} test application (\testap{}) backed by a runtime \ac{UEFI} environment.
We fuzz \testap{} using a coverage-guided fuzzer and apply \ac{LLM}-assisted crash triaging to filter false positives.

\subsection{Overview}
\label{sec:overview}

We use \lst{lst:newrunningexample} to provide an overview of our technique, which shows the DHCP driver (\texttt{Dhcp4Dxe}) of \edk{}, in which \systemname{} discovered two real vulnerabilities.
\tobepublished{}

\systemname{} operates in two phases: \emph{Rehosting} and \emph{Testing}, given a target protocol \targetprotocol{} and the \edk{} source code.
The rehosting phase consists of several steps, all of which have access to the target protocol and the full \edk{} source code.
First, in the \emph{Dependency Identification} step (\sect{subsubsec:depidentification}), we identify all dependencies of \targetprotocol{}, \ie{} \dependenciesforp{\targetprotocol{}}. 
Static dependencies (\eg{} libraries) are extracted from build files, while protocol dependencies (\ie{} those accessed through \texttt{OpenProtocol} service) require source-level analysis, which checks standard protocol API calls (\ie{} \texttt{LocateProtocol}) and records the GUID associated with call that identifies the protocol.
For the example in \lst{lst:newrunningexample}, we obtain library dependencies from build configurations and analyze the source code to identify protocol interactions. 

\begin{wrapfigure}[40]{R}{0.5\textwidth}
\vspace{-\baselineskip}
\begin{minted}[xleftmargin=0.15cm, numbersep=1pt, fontsize=\tiny, linenos, breaklines, escapeinside=||]{cpp}
|\colorbox{green!25}{\faCogs}| EFI_STATUS EFIAPI InitializeHarness (VOID) {
  EFI_STATUS  Status;
  // Initialize target protocols driver.
  Status = HostDispatchDrivers (gFuzzHandle, gHostDriverRegistry);
  ...
  // get the handle for target protocol
  Status = gBS->HandleProtocol (gFuzzHandle,
                &gEfiDhcp4ServiceBindingProtocolGuid, ...);
  ...
}

|\colorbox{green!25}{\faCogs}| VOID EFIAPI RunTestHarness (IN MOCK_FUZZ_CONTEXT  *FuzzCtx) {
  |\colorbox{yellow!40}{\faRobot}| // start
  ApiSelector = |\colorbox{green!25}{\faArrowCircleLeft}| MockFuzzContextGetU8 (FuzzCtx);
  ...
  // invoke interface dependencies
  mProtocol->Configure (mProtocol, &PreCfg);
   // API dispatch
  |\colorbox{green!25}{\faCogs}| switch (ApiSelector % API_COUNT) {
    ...
    case API_TRANSMITRECEIVE:
      // fuzz target interface function.
      FuzzTransmitReceive (FuzzCtx, SubOptions);
      break;
    case API_PARSE:
      FuzzParse (FuzzCtx, SubOptions);
    ...
  }
  |\colorbox{yellow!40}{\faRobot}| // end
}

|\colorbox{green!25}{\faCogs}| STATIC VOID FuzzParse (IN MOCK_FUZZ_CONTEXT  *FuzzCtx,
        IN UINT8  SubOptions) {
  |\colorbox{yellow!40}{\faRobot}| // start
  EFI_DHCP4_PACKET LocalPacket;
  ...
  LocalPacket.Size   = sizeof (LocalPacket);
  UINT8 *_p = (UINT8 *)&LocalPacket.Dhcp4;
  UINTN  _cap = LocalPacket.Size - OFFSET_OF (EFI_DHCP4_PACKET, Dhcp4);
  UINTN _fuzz_max = _cap - 1;
  UINT8 *_src = |\colorbox{green!25}{\faArrowCircleLeft}| MockFuzzContextConsume (FuzzCtx, _fuzz_max);
  if (_src != NULL) { CopyMem (_p + _fuzz_off, _src, _fuzz_max); }
  _p[_cap - 1] = 0xFF;
  LocalPacket.Length = (UINT32) _cap;
  ...
  mProtocol->Parse (NULL, &LocalPacket, ...);
  |\colorbox{yellow!40}{\faRobot}| // end
}
\end{minted}
\vspace{-0.5\baselineskip}
\captionof{listing}{Harness for the DHCP protocol generated by \systemname{}, where \colorbox{green!25}{\faCogs{}} and \colorbox{yellow!40}{\faRobot} indicate content was generated through templates and \ac{LLM}, respectively. \colorbox{green!25}{\faArrowCircleLeft} indicates the {\tt FuzzContext} API calls.}
\label{lst:dhcpharness}
\vspace{-\baselineskip}
\end{wrapfigure}

Second, using the identified dependencies, we generate harnesses and mocks in parallel.
Both \emph{Mock Generation} (\sect{subsubsec:mockgeneration}) and \emph{Harness Generation} (\sect{subsubsec:harnessgeneration}) steps use a template-guided approach with \ac{LLM} assistance to generate semantically equivalent mocks and effective harnesses, respectively.
Specifically, the generated harness \ie{} \harnessforprotocol{\targetprotocol{}}, invokes protocol interface functions (\ie{} \protocolfunctionset{\targetprotcol{}}) by translating fuzzer input into well-typed arguments.
\lst{lst:dhcpharness} shows the harness generated for DHCP (our running example in \lst{lst:newrunningexample}).
Similarly, the generated mock uses fuzzer input to provide the output (\ie{} data required by \targetprotocol{}), \lst{lst:lowfidelitymock} and \ref{lst:udpcomplexstruct} show snippets of functions in the mock generated for UDP, \ie{} a direct dependency of DHCP.
We also identify external interaction points and replace them with input-driven stubs that derive their behavior from fuzzer inputs.
We generate mocks for all the direct dependencies of \targetprotocol{}, \ie{} \dependenciesforpatlevel{1}{\targetprotocol{}}

Third, we assemble (\sect{subsubsec:usermodeassemble}) the generated harness, mocks, and the target protocol with a runtime library that provides the necessary \ac{UEFI} environment (\ie{} \inlinecode{MetisLib}), producing the user-space \edk{} test application for \targetprotocol{} (\ie{} \testapproto{\targetprotocol{}}).
We also use a dynamic refinement technique (\sect{subsubsec:dynamicrefinement}) to improve the generated harnesses and mocks by small test runs.

In the testing phase, we fuzz the final \testapproto{\targetprotocol{}} using \afl{}.
During this phase, we utilize a 5 round refinement loop step used to iteratively test, collect coverage metrics, improve the harnesses and mocks, and then repeat, which aids in faulty harnesses and mocks to reach greater code coverage as seen in our evaluation (\sect{subsubsec:effectresults}).
Since some crashes may arise due to imprecise mocks, we employ a validation step (\sect{subsubsec:crashtriaging}) using context-guided filtering to identify false positives, thereby isolating true vulnerabilities.





\subsection{Phase 1: Rehosting}
Our goal in this phase is to generate \testapproto{\targetprotocol{}}, a self-contained user-space program in which every input channel to \targetprotocol{} is either exposed through a real protocol interface or controlled by fuzzer input via the generated mocks for \dependenciesforp{\targetprotocol{}}.
This phase has the following stages:

\subsubsection{Dependency Identification}
\label{subsubsec:depidentification}
For a given protocol \targetprotocol{} the goal of this step is to extract all its direct dependencies (\dependenciesforp{\targetprotocol{}}), their interface structures, and interface dependencies (\interfacedeps{\targetprotocol}).
This analysis operates at the component level: as discussed in \sect{subsubsec:notationsanddefinitions}, a protocol inherits all dependencies of the component that implements it.

\noindent\textbf{Protocol Dependencies.}
In \edk{}, each component provides an {\tt INF} file that declares the protocols it consumes and produces using {\tt CONSUMES} and {\tt PRODUCES} keywords, respectively.
We consider the consumed protocols as \emph{explicit dependencies}.
We identify these by parsing the {\tt INF} file to extract the protocols listed under {\tt CONSUMES} tag.
For the running example (\lst{lst:newrunningexample}), we identified \inlinecode{gEfiUdp4ProtocolGuid} and \inlinecode{gEfiUdp4ServiceBindingProtocolGuid} as explicit dependencies through the {\tt INF} file.


The information in the {\tt INF} file is often incomplete, as dependencies introduced through library code are not explicitly listed; we call these \emph{implicit dependencies}.
For example, the DHCP driver (\lst{lst:newrunningexample}) invokes \inlinecode{PseudoRandomU32} a library function (of {\tt DxeNetLib} library) that internally uses \inlinecode{RNG_PROTOCOL} protocol, which does not appear in the driver's {\tt INF} file.
We perform source-level analysis to identify implicit dependencies.
Specifically, we detect calls to \inlinecode{LocateProtocol} to identify protocol usage, and recursively analyze library calls (\eg{}  \inlinecode{PseudoRandomU32}) to uncover additional protocol dependencies.
For the running example (\lst{lst:newrunningexample}), we identified \inlinecode{EFI_RNG_PROTOCOL} as an implicit dependency, consumed transitively through \inlinecode{DxeNetLib}'s \inlinecode{PseudoRandomU32} function.
We use the dependencies and the information from {\tt PRODUCES} tag to identify dependencies between the given component and other component.
\extonly{\Apdxref{\fig{fig:working-example}} shows the partial snippet of the dependency information extracted for the DHCP driver (\ie{} DHCP Protocol).}

\noindent\textbf{Interface Dependencies.}
\ac{UEFI} protocols maintain state either through global variables or protocol handle objects created during \inlinecode{OpenProtocol} calls. \eg{} \inlinecode{UdpIo->UdpHandle} in the DHCP driver.
These handle objects encapsulate protocol-specific state via a set of data fields that are read and updated by different interface functions, thereby inducing dependencies between them.

To capture these dependencies, we first identify all functions that define or update each field in the handle object.
Next, for each interface function $f$, we analyze whether its execution is \emph{exit control dependent} on the values of any handle fields, \ie{} a control dependency (\sect{subsubsec:depidentification}) if not satisfied results in premature return from the function.
For each such field, we identify the interface functions that initialize or modify it, and record them as interface dependencies of $f$, \ie{} add them to \interfacedepsfunc{\targetprotocol}{$f$}. 
For instance, the following control dependency:
\inlinecode{if (!Instance->Configured)} \inlinecode{return EFI_NOT_STARTED}
marks the function to be dependent on the \inlinecode{Configured} field, and any interface function writing or modifying \inlinecode{Configured} will be considered as its dependent, \ie{} will be added to \interfacedepsfunc{\targetprotocol}{$f$}.
We apply the same analysis to global variables to capture dependencies induced by shared state.
Our analysis is interprocedural, allowing us to capture dependencies across function boundaries and recover the complete set of interface dependencies. Moreover, we follow firmware initialization sequences (such as DriverEntry and protocol installation order) to ensure that the memory allocation and initialization order is respected.

\subsubsection{Mock Generation}
\label{subsubsec:mockgeneration}
The goal of mock generation is to create a \emph{MockProtocol} for each dependency of \targetprotocol{}, \ie{} \dependenciesforpatlevel{1}{\targetprotocol{}}.
For a given protocol $x \in \dependenciesforpatlevel{1}{\targetprotocol{}}$, the corresponding MockProtocol implements the same interface as $x$, with behavior driven by fuzzer-provided input.
Specifically, for each interface function of $x$ (\ie{} \protocolfunctionset{x}), we focus on generating well-formed outputs using input data.

\begin{wrapfigure}{R}{0.4\textwidth}
\centering
\includegraphics[width=0.4\textwidth]{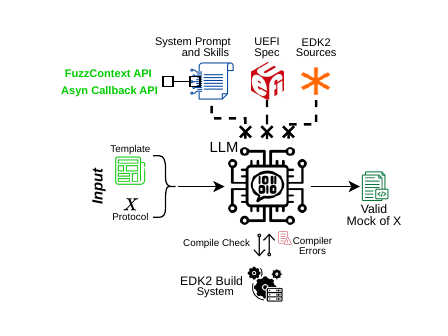}
\caption{\ac{LLM}-Assisted Mock Generation Technique used by \systemname{}}
\label{fig:mockgeneration}
\end{wrapfigure}

We generate mocks using a template-driven approach, augmented with \ac{LLM}-assisted synthesis.
\fig{fig:mockgeneration} shows the overview of our technique.

\noindent\emph{Template Generation.}
For a given protocol, we generate a mock template containing one stub for each interface method.
Using the interface structure recovered during dependency identification, we extract the signature of every function and create a corresponding stub with the same signature.
For example, for the UDP protocol required by the DHCP protocol in \lst{lst:newrunningexample}, we use the interface structure (\ie{} {\tt EFI\_UDP4\_PROTOCOL}\extonly{ in \apdxref{\lst{lst:udpheader}}}) to identify all interface methods.
For each method (\eg{} {\tt GetModeData}), we use its type declaration (\ie{} {\tt EFI\_UDP4\_GET\_MODE\_DATA}\extonly{ in \lst{lst:udpheader}}) to generate a matching stub function (\ie{} {\tt MockUdp4GetModeData}\extonly{, lines 16-20 in \lst{lst:lowfidelitymock}}).
We additionally generate a constructor routine that installs the mock protocol by initializing the interface structure’s function pointers with the generated stub functions.
\extonly{For instance, lines 7-13 in \lst{lst:lowfidelitymock} show the constructor generated for the UDP protocol.}

\noindent\emph{Interface Method Generation.}
Each stub is responsible for producing well-formed outputs by updating output parameters and return values using fuzzer-provided data.
A naive approach would treat all pointer parameters as outputs; however, this is both inefficient and incorrect.
Instead, we use \edk{}'s \inlinecode{OUT} and \inlinecode{INOUT} annotations to identify output parameters.
Specifically, \edk{} coding standards require all parameters to be explicitly annotated as input (\inlinecode{IN}), output (\inlinecode{OUT}), or input and output (\inlinecode{INOUT})~\cite{uefi-spec}.

A straightforward approach is to populate output parameters using raw bytes from the input stream, but this often violates type invariants (\eg{} for pointer fields), leading to spurious crashes\extonly{ \eg{} {\tt EFI\_UDP4\_FRAGMENT\_DATA} in \apdxref{\lst{lst:complexstruct}}}.
A more structured alternative is type-aware generation~\cite{muralee2025reactivebottomuptesting}, which recursively constructs values based on their types.
However, this approach still fails to capture semantic constraints such as valid value ranges or inter-field relationships (\eg{} size–buffer consistency), which are critical for meaningful execution.
For instance, the {\tt DataLength} field\extonly{ in \lst{lst:complexstruct}} should be the total length of all fragments, \ie{} sum of {\tt FragmentLength} field of all entries in {\tt FragmentTable} field.
Similarly, {\tt FragmentLength} should be the number of entries pointed by the {\tt FragmentTable} field.

We observe that the type constraints are often documented in \ac{UEFI} specification~\cite{uefi-spec}, and source code comments\extonly{, \apdxref{\lst{lst:udpuefspec}} shows a snippet from the UEFI specification describing the semantics of different fields of the {\tt EFI\_UDP4\_RECEIVE\_DATA} structure}.
\extonly{Also, the source code of \edk{} contains references to {\tt BOOTREPLY}, \ie{} constant value {\tt 0x63825363} present in the DHCP packet indicating a response for a network boot request.
These values could also be used by \ac{LLM} to generate valid network packets.
}
Recent advances in \acp{LLM} have shown strong capability in extracting such semantic information and generating correct code ~\cite{yang2025harnessagentscalingautomaticfuzzing, Zhang_2024, PromeFuzz, Liu_OSS_Fuzz_Gen_Automated_Fuzz_2024}.
Motivated by these observations, we employ an \ac{LLM}-assisted approach.
\fig{fig:mockgeneration} shows the overview of our technique.

We want the \ac{LLM} to generate code that correctly handles all outputs of an interface function.
In addition to explicit output parameters (\ie{} annotated with {\tt OUT}), interface functions in \edk{} also communicate through events and callbacks, as discussed in \sect{subsec:edk2}.
These callbacks are associated with events and are triggered when the corresponding signal is raised (\eg{} {\tt RaiseTPL}).

\noindent\emph{Handling Events/Callbacks.}
However, \edk{} event callbacks are platform-dependent, rely on hardware support, and cannot be directly used in user space.
To address this, we introduce the {\tt AsyncCallBack} API, which emulates the event and callback semantics of \edk{}.
\extonly{\Apdxref{\tbl{tab:asynccallbacklib}} summarizes the API methods.}
We provide the \ac{LLM} with the API description and examples demonstrating how to model different \edk{} callback patterns using {\tt AsyncCallBack} API.

\noindent\emph{Streamlined Input Consumption.}
Interface functions consume fuzzing input in diverse ways, making input handling difficult to generalize.
To simplify this process, we design \inputlib{}\extonly{ (shown in \apdxref{\tbl{tab:fuzzcontexlib}})}, a utility library that provides a uniform interface for consuming fuzzing input and constructing required data structures.
We provide the \ac{LLM} with both the library description and usage examples.

\noindent\emph{Handling Constants and Return Codes.}
When multiple valid constant values exist (\eg{} structure fields), we instruct the \ac{LLM} to select among them using fuzzing input\extonly{, \eg{} line 21 in \lst{lst:udpcomplexstruct}}.
Similarly, to improve coverage of error-handling paths, we instruct the \ac{LLM} to occasionally return failure codes with low probability\extonly{, \eg{} lines 33-35 in \lst{lst:lowfidelitymock}}.

In summary, we incorporate the descriptions of the {\tt AsyncCall} API, \inputlib{}, and representative examples into the system prompt.
As shown in \fig{fig:mockgeneration}, We additionally provide access to \ac{UEFI}/\edk{} documentation and source code through a curated knowledge base included in each prompt and a clangd-backed Serena~\cite{serena} MCP server for symbol navigation and file search.
We also provide access to the \edk{} build system to guide the \ac{LLM} in producing syntactically valid code.
\extonly{\Apdxref{Prompt~\ref{pr:mock-generation-prompt}} shows a snippet of the mock generation prompt used by \systemname{}.
\lst{lst:udpcomplexstruct} shows the snippet of the code generated for {\tt MockUdp4Recieve} interface function, it shows how the constant values (lines 17, 19) from \edk{} documentation are used to create valid DHCP packets, and also the use of {\tt AsyncCallBack} (\faClock[regular]) and {\tt \inputlib{}} (\faArrowCircleLeft).}

\begin{listing}[t]
\begin{minipage}[t]{0.49\textwidth}
\begin{minted}[xleftmargin=0.15cm, numbersep=1pt, fontsize=\tiny, linenos, breaklines, escapeinside=||]{cpp}
|\colorbox{green!25}{\faCogs}| STATIC EFI_UDP4_PROTOCOL mMockUdp4Protocol = {
  MockUdp4GetModeData,
  MockUdp4Configure,
  ...
};

|\colorbox{green!25}{\faCogs}| MockUdp4Constructor (VOID) {
 ...
 Status = gBS->InstallProtocolInterface (&gFuzzHandle,
              &gEfiUdp4ProtocolGuid, EFI_NATIVE_INTERFACE,
              &mMockUdp4Protocol);
 ...
}

|\colorbox{green!25}{\faCogs{}}| STATIC EFI_STATUS EFIAPI
MockUdp4GetModeData (
  IN EFI_UDP4_PROTOCOL* This,
  OUT EFI_UDP4_CONFIG_DATA* Udp4ConfigData,
  OUT EFI_IP4_MODE_DATA* Ip4ModeData,
  ...) {
  |\colorbox{yellow!40}{\faRobot}| // Start
  MOCK_FUZZ_CONTEXT  *FuzzCtx = |\colorbox{green!25}{\faArrowCircleLeft}| ResolveContext ();
  ...
  if (Udp4ConfigData != NULL) {
    ZeroMem (Udp4ConfigData, sizeof (EFI_UDP4_CONFIG_DATA));
    if (FuzzCtx != NULL) {
      UINT8 *Bytes = |\colorbox{green!25}{\faArrowCircleLeft}| MockFuzzContextConsume (FuzzCtx, sizeof (EFI_UDP4_CONFIG_DATA));
      if (Bytes != NULL) { CopyMem (Udp4ConfigData, Bytes, sizeof (EFI_UDP4_CONFIG_DATA)); }
    }
  }
  ...
  if (FuzzCtx != NULL) {
    UINT8 Choice = |\colorbox{green!25}{\faArrowCircleLeft}| MockFuzzContextGetU8 (FuzzCtx);
    if (Choice == 0xFF) { return EFI_NOT_READY; }
    if (Choice == 0xFE) { return EFI_OUT_OF_RESOURCES; }
  }
  return EFI_SUCCESS;
  |\colorbox{yellow!40}{\faRobot}| // End
}
  
\end{minted}
\caption{Mock generated for UDP, where \colorbox{green!25}{\faCogs{}} and \colorbox{yellow!40}{\faRobot} indicate content was generated through templates and \ac{LLM}, respectively. \colorbox{green!25}{\faArrowCircleLeft} indicates the {\tt FuzzContext} API calls.}
\label{lst:lowfidelitymock}

\end{minipage}\hfill
\begin{minipage}[t]{0.49\textwidth}
\begin{minted}[xleftmargin=0.15cm, numbersep=1pt, fontsize=\tiny, linenos, breaklines, escapeinside=||]{cpp}
MockUdp4Receive ( IN EFI_UDP4_PROTOCOL* This,
  IN EFI_UDP4_COMPLETION_TOKEN* Token)
{
  MOCK_FUZZ_CONTEXT  *FuzzCtx = |\colorbox{green!25}{\faArrowCircleLeft}| ResolveContext ();
  ...
  if (FuzzCtx != NULL && Token != NULL) {
    UINTN PktLen = 576;
    UINT8 *PktSrc = |\colorbox{green!25}{\faArrowCircleLeft}| MockFuzzContextConsume (FuzzCtx, PktLen);
    if (PktSrc != NULL) {
      EFI_UDP4_RECEIVE_DATA *RxData = AllocateZeroPool (sizeof (EFI_UDP4_RECEIVE_DATA) + PktLen);
      if (RxData != NULL) {
        // Copy input data
        CopyMem (RxData + 1, PktSrc, PktLen);
        ...
        // Fix ups from UEFI Spec
        UINT8 *Pkt = (UINT8 *)(RxData + 1);
        Pkt[0] = 2; Pkt[1] = 1; Pkt[2] = 6;
        if (PktLen > 244) {
          Pkt[236] = 0x63; Pkt[237] = 0x82; Pkt[238] = 0x53; Pkt[239] = 0x63;
          ...
          Pkt[242] = (|\colorbox{green!25}{\faArrowCircleLeft}| MockFuzzContextGetU8(FuzzCtx) % 2 == 0) ? 2 : 5;  
          Pkt[243] = 0xFF;
        }
      }
      RxData->DataLength = (UINT32)PktLen; |\scalebox{0.6}{\faLink}$_{l}$|
      RxData->FragmentCount = 1;
      RxData->FragmentTable[0].FragmentLength = (UINT32)PktLen; |\scalebox{0.6}{\faLink}$_{l}$|
      RxData->FragmentTable[0].FragmentBuffer = (UINT8 *)(RxData + 1); |\scalebox{0.6}{\faLink}$_{b}$|
      Token->Packet.RxData = RxData;
     }
   } 
  ...
  if (Token != NULL && Token->Event != NULL) {
    Token->Status = EFI_SUCCESS;
    |\colorbox{green!25}{\faClock[regular]}| MockDeferredSchedule (&mMockUdp4DeferredState, Token->Event);
  }
 ...
} 
\end{minted}
\caption{Snippet of \systemname{} mock illustrating handling of buffer/length correlations (\scalebox{0.6}{\faLink}$_{l}$, \scalebox{0.6}{\faLink}$_{b}$), constant value fix-ups, and event call-backs using our {\tt AsyncCallBack} API (\colorbox{green!25}{\faClock[regular]}).}
\label{lst:udpcomplexstruct}

\end{minipage}
\end{listing}

\subsubsection{Harness Generation}
\label{subsubsec:harnessgeneration}

This step focuses on generating a testing harness for the target protocol \targetprotocol{}.
Similar to mock generation, we adopt a template-based approach augmented with \ac{LLM} assistance.
Each harness consists of (i) initialization code and (ii) invocations of interface functions.
\lst{lst:dhcpharness} shows the snippet of the harness generated by \systemname{} for DHCP protocol (in \lst{lst:newrunningexample}).

\noindent\emph{Initialization Code.}
We generate initialization code using fixed templates. These templates capture common patterns for initializing a protocol’s component.
For example, if \targetprotocol{} is implemented by a driver, we insert calls to the corresponding driver initialization routines (\eg{} \inlinecode{DriverEntryPoint}, \inlinecode{DriverBinding.Start}).

We also obtain a handle to \targetprotocol{} using \inlinecode{LocateProtocol} bootservice. We generate the variables required for initialization using interface structure types identified in the Dependency Identification (\sect{subsubsec:depidentification}) step.
The {\tt InitializeHarness} function in \lst{lst:dhcpharness} shows the snippet of the initialization code generated by \systemname{}.



\noindent\emph{Invoking Interface Functions.}
Next, we generate calls to each interface function of \targetprotocol{} (\ie{} \protocolfunctionset{\targetprotcol{}}).
Specifically, we read a byte from the input and use its value to select and invoke one of the interface functions.
As shown in \lst{lst:dhcpharness}, we read a byte ({\tt ApiSelector}) and use it to select the interface function (line 19).


Before invoking a function $f \in$\protocolfunctionset{\targetprotcol{}}, we ensure that all its interface dependencies (\ie{} \interfacedepsfunc{\targetprotocol}{f}) are satisfied by invoking the required predecessor functions, \ie{} we invoke all functions in \interfacedepsfunc{\targetprotocol}{f} before invoking $f$. \eg{} In \lst{lst:dhcpharness}, {\tt Configure} is invoked (at line 17) before {\tt receive} (at line 23).

\noindent\emph{Argument Generation.}
For each interface function invocation, we generate values for its input arguments (\ie{} parameters annotated with \inlinecode{IN} or \inlinecode{INOUT}) using an \ac{LLM}-assisted approach (described in \sect{subsubsec:mockgeneration}).

A key challenge arises with \emph{state-dependent} types, \ie{} the type depends on the protocol internal state or prior interactions.
In \edk{}, there are many such types that are expected to be produced and initialized by earlier interface calls.
Constructing such structures directly from raw input data often leads to invalid states and immediate failures.

To address this, we use the idea of \emph{generator functions}~\cite{glosner_fuzzuer_2025}.
For a given type $t$, we identify all interface functions that produce $t$ as an \inlinecode{OUT} parameter and treat them as generators for $t$.
We then use these generators to construct valid instances of $t$ by invoking the corresponding interface functions before use. Specifically, we provide this generator information to the \ac{LLM}, enabling it to synthesize argument construction code that respects protocol state and execution ordering.


\noindent\emph{Handling Complex Semantics.}
Certain protocols (\eg{} network interactions) impose additional semantic constraints, such as required invocation sequences (\eg{} \inlinecode{Send} before \inlinecode{Receive}).
To handle such cases, we incorporate domain-specific knowledge and patterns into the \ac{LLM} prompts.

\extonly{\Apdxref{Prompt~\ref{pr:harnessgenerationprompt}} shows a snippet of the harness generation prompt used by \systemname{}.}
Additionally, similar to our mock generation approach, we also provide the \ac{LLM} with access to \edk{} source code and documentation via Serena~\cite{serena} MCP server for symbol navigation and file search, allowing it to incorporate implicit constraints and usage patterns when generating inputs.

\subsubsection{User Mode Assemble}
\label{subsubsec:usermodeassemble}
This stage combines the harness, mocks (M(\dependenciesforpatlevel{1}{\targetprotocol{}})), the target protocol (\targetprotocol{}), and our runtime library (\metislib{}) into a statically linked user-space ELF binary \testapproto{\targetprotocol{}}, instrumented with \afl{} and \ac{ASAN}/\ac{UBSan}.

\metislib{} provides semantically equivalent implementations of \ac{UEFI} Boot Services and \edk{} library classes using Linux user-space APIs.
We develop \metislib{} as a one-time effort to enable seamless rehosting of \edk{} components in a user-space environment.

We also generate necessary build artifacts, \ie{} {\tt DSC} and {\tt INF} files, required by the \edk{} build system to build the \testapproto{\targetprotocol{}} by linking with the mocks, harness, and \metislib{}.\extonly{ This enables us to reuse \edk{} build system and generate the \testapproto{\targetprotocol{}} binary by using standard build command, \ie{} \inlinecode{build -a X64 -t AFL -b DEBUG}.}

\subsubsection{Dynamic Refinement}
\label{subsubsec:dynamicrefinement}
In this step, we iteratively refine the harness and mocks to fix incomplete setup and inaccurate mock behavior.
We perform up to five refinement rounds, where each round consists of fuzzing for 2 minutes followed by coverage collection.
We analyze the coverage to identify \emph{blocking frontiers}, \ie{} conditional statements that guard currently uncovered code.
An \ac{LLM} agent attributes each blocking frontier to one of the following causes and revises the mocks, the harness, or the seed corpus accordingly.

\begin{itemize}[leftmargin=*, noitemsep]
\item \textbf{Missing Initializations.}
The driver never completes \inlinecode{Start} because a mock does not provide what it requires\extonly{ (\eg{} a service binding whose \inlinecode{CreateChild} returns a handle without the child protocol)}, leaving all later code unreachable.
We fix the mock behind the first failing call\extonly{ before addressing any other cause}.
\item \textbf{Data Deficiency.}
The condition enforces constraints that the current mocks or inputs do not satisfy\extonly{: a mock returns an error status where the real protocol would succeed or an implausible structure (\eg{} a zero \inlinecode{BlockSize}), or the input is shorter than the header the driver parses}.
We update the mocks, the harness, or the seed corpus so that these constraints hold (\eg{} lines 14-29 in \lst{lst:dhcpharness}), without making mocks succeed unconditionally.
\item \textbf{Missing Completions.}
An asynchronous mock accepts a request but never signals its completion event, so all completion paths remain uncovered.
We revise the mock to schedule the completion and the harness to advance simulated time.
\end{itemize}

Testing-unfriendly patterns such as hardware waits and polling loops need no refinement, as our host libraries already bound them (\eg{} \inlinecode{Stall} is a no-op and \inlinecode{CheckEvent} advances simulated time).
After each round, we recreate the \testapproto{\targetprotocol{}} (\sect{subsubsec:usermodeassemble}) using the revised harness and mocks and fuzz it again; we keep a revision only if it covers at least four more driver lines than the best version so far, and otherwise restore that version and stop.
The best \testapproto{\targetprotocol{}} after at most five rounds is used in the next testing phase.

\subsection{Phase II: Testing}
Given the final \testapproto{\targetprotocol{}}, this phase aims to discover vulnerabilities and distinguish real bugs from artifacts introduced by the mocking layer.
It consists of two concurrent steps.
\extonly{The fuzzing step (\sect{subsubsec:fuzzingstep}) runs continuously, exploring new inputs and triggering potential crashes.
Whenever a crash is detected, the crash triaging step (\sect{subsubsec:crashtriaging}) is invoked to analyze the failure and filter out false positives.}

\subsubsection{Fuzzing}
\label{subsubsec:fuzzingstep}
We fuzz \testapproto{\targetprotocol{}} using \afl{} in persistent mode~\cite{fioraldi2020afl}, seeded with \afl{}’s default inputs and the corpus generated during refinement.
In each iteration, the harness may invoke multiple interface functions based on the fuzzer input.
To streamline input delivery, we use a shared context structure to store the fuzzer input.
Both the harness and mocks access input data through this structure.
This unified design ensures consistent input handling and enables deterministic execution across fuzzing iterations.

\subsubsection{Crash Triaging}
\label{subsubsec:crashtriaging}
The primary goal of triaging is to eliminate false positives and retain only true vulnerabilities.
In particular, \emph{we treat crashes that occur within mock protocols or the harness as false positives}, since they may arise from our approximations rather than genuine bugs in \edk{}.

For each crash in \targetprotocol{}, we extract a fingerprint and use it for deduplication and analysis.
Specifically, we re-execute \testapproto{\targetprotocol{}} with the crashing input and extract a fingerprint of the form \texttt{\textit{type}\textbar{}\allowbreak\textit{function}:\allowbreak\textit{file}:\allowbreak\textit{line}} from the top stack frame in the ASAN/UBSAN report.
We then deduplicate crashes based on this fingerprint.

For each remaining crash, we analyze whether it occurs while processing attacker-controlled input (\eg{} network packets, PE-COFF headers, or image buffers).
Such crashes are classified as \emph{REAL\_BUG}, while the remaining cases are labeled \emph{NEEDS\_REVIEW} for manual inspection.
We use \ac{LLM} assistance to perform this filtering. Specifically, we provide access to the crashing input, \testapproto{\targetprotocol{}}, and {\tt gdb} commands. Our prompt includes various examples and process to triage a crash.\extonly{ The snippet of the prompt is shown in \apdxref{Prompt~\ref{pr:crashtriagingprompt}}.}

\subsection{Implementation}
\label{sec:implementation}

The core components of \systemname{}, including dependency identification and static analyses for extracting information required for mock and harness generation, are implemented in python.
We use \texttt{openrouter agent sdk}~\cite{openrouter_agent_sdk} to implement our \ac{LLM}-assisted techniques.
All \ac{LLM}-generated code is validated for syntactic correctness via compilation checks, and any failures are rejected.
\subsubsection{Runtime Library}
Our runtime library, \metislib{}, is implemented in C (27,867 SLoC).
\metislib{} is a self-contained host runtime that re-implements the \edk{} \inlinecode{LibraryClass} ABI using POSIX APIs, enabling unmodified driver code to link and execute as a user-space ELF binary.
It provides host-side implementations for commonly used library classes (\eg{}Base, Memory, BootServices, RuntimeServices, DXE, HII, PCD, SMM), while unsupported classes fall back to \edk{}’s existing null stubs.
A key component is the host implementation of \inlinecode{UefiBootServicesTableLib}, which maintains an in-memory handle database that preserves the ordering semantics of \inlinecode{InstallProtocolInterface}, \inlinecode{OpenProtocol}, and \inlinecode{LocateProtocol}.
This allows drivers to interact transparently with \inlinecode{gBS}, \inlinecode{gRT}, and \inlinecode{gDS} without modification.A user-space dispatcher iterates over the harness-generated driver registry and invokes each driver’s entry point and \inlinecode{DriverBinding.Supported}/\inlinecode{Start} routines using synthetic controller handles, mirroring the behavior of platform BDS after device enumeration.


\subsubsection{Helper Libraries} Our helper libraries, \ie{} \inputlib{} and {\tt AsyncCall} API\extonly{ (see \apdxref{Tables~\ref{tab:fuzzcontexlib} and~\ref{tab:asynccallbacklib}})}, are implemented in C (916 SLoC).
\extonly{As mentioned in \sect{subsubsec:mockgeneration}, these libraries assist \ac{LLM} in streamlining the generation of mocks and harnesses.}
\section{Evaluation}
\label{sec:evaluation}
We structure our evaluation along the following research questions:

\begin{itemize}[noitemsep]
    \item \textbf{RQ1 (Rehosting Effectiveness):}
    How many \edk{} drivers can \systemname{} successfully rehost?
    How well do the generated harnesses exercise the protocol surface, and how faithfully do the generated mocks model dependency semantics?

  \item \textbf{RQ2 (Fuzzing Effectiveness):}
   How effective is \systemname{}'s approach at testing \edk{} drivers? What is the code coverage and effectiveness at identifying vulnerabilities?

  \item \textbf{RQ3 (Comparative Evaluation):}
    How does \systemname compare to existing techniques?


    \extonly{\item \textbf{RQ4 (Generalizability and Practicality):}
    We report the RQ4 evaluation in \apdxref{\sect{sec:eval-rq4-practicality}}.}


\end{itemize}

\subsection{Setup}
\label{sec:eval-setup}
\extonly{We used the following setup for our evaluation.}
\subsubsection{Target selection.}
\label{subsubsec:targetselection}
We evaluate \systemname{} on \numdrivers{} drivers from the mainline \edk{} repository (\texttt{f0542ae07d}), spanning six subsystem categories: Network (22), PCI/Storage (16), USB (7), Console/HII (9), SMM/Firmware (4), and Utility (2).
A single \edk{} driver can produce multiple protocols; together, these \numdrivers{} drivers produce the \numprotocols{} protocols we evaluate.
The complete list of protocols, grouped by category, is provided in \tbl{tab:fuzzingeval}.
\tbl{tab:fuzzingeval} also shows the total number of interface functions ($\mid\protocolfunctionset{\targetprotcol{}}\mid$) in each protocol, functions with at least one complex \inlinecode{struct *} argument ($ComF$), and functions with at least one \inlinecode{void *} argument ($VoidF$).\extonly{
These protocols exhibit significant diversity in the number and types of interface functions, as well as in the complexity of their dependency structures.}


\subsubsection{Known Vulnerabilities.}
We curated \numpreviousbugs{} known vulnerabilities
, selected based on their public CVE entries and diversity across drivers and bug classes.
These were injected into our \edk{} commit (\inlinecode{f0542ae07d}) by porting the vulnerable pre-patch code back on top of the patched tree\extonly{, isolated to per-driver overlays so the rest of the codebase is unchanged}.

\subsubsection{\ac{LLM} choices}
\label{subsubsec:llmchoices}
\systemname{} implementation is \ac{LLM}-agnostic and can be used with any \ac{LLM}.
We evaluate the pipeline using six \acp{LLM}\extonly{ (see \apdxref{\tbl{tab:cov_per_llm}}) spanning frontier models, such as Claude Sonnet~5.5,  and also open-weight models (\eg{} DeepSeek)}.
Unless otherwise specified, \emph{we use \textbf{Claude Sonnet~5.5} as the default model, as it achieved the highest aggregate coverage and the fewest compilation errors\extonly{ (\apdxref{\sect{sec:eval-rq4-practicality}})}}.


\subsubsection{Fuzzing Setup.}
\extonly{
All experiments are run on an AWS EC2 \texttt{c5.metal} instance (Intel Xeon server, 96 cores, 188\,GB RAM) with Ubuntu~24.04 as the host OS, inside a Docker container running Ubuntu~22.04.}
Each target is fuzzed for 24\,h on a single AFL++~v4.33c instance pinned to one core, with ASan and UBSan enabled — matching the per-target budget of all baselines and the guidelines of Klees~\etal~\cite{klees2018evaluating}.


\subsection{RQ1: Rehosting Effectiveness}
\label{sec:eval-rq1}
\extonly{We want to assess the success rate of \systemname{} in rehosting \edk{} drivers and the accuracy of the generated harnesses and dependency mocks.}
\subsubsection{Method}
\label{subsubsec:evalrehostingmethod}
To measure rehosting success, we report the number of drivers for which our pipeline successfully produces a compilable and executable \testapproto{\targetprotocol{}}.

\noindent\emph{Harness Accuracy.}\extonly{
A harness is effective if it enables meaningful testing of all interface functions of the target protocol \targetprotocol{}.}
We assess harness accuracy using the following metrics:
\begin{itemize}[leftmargin=*, noitemsep]
\item \emph{Interface Completeness (Com).} Fraction of interface functions that are invoked by the harness.
\item \emph{Argument Completeness (Arg).} Fraction of interface functions that are invoked with correct arguments.
\item \emph{Dependency Handling (DepH).} Fraction of interface functions with dependencies (\ie{} $\forall f \in \protocolfunctionset{\targetprotcol{}} \mid \interfacedepsfunc{\targetprotcol{}}{f} \neq \emptyset{}$) whose dependencies are correctly satisfied.
\item \emph{Handling Complex Arguments (CArg).} Fraction of stateful or complex arguments that are correctly constructed.
\end{itemize}

\noindent\emph{Mock Accuracy.}\extonly{
Ideally, generated mocks should faithfully mimic the behavior of the corresponding protocols.}
We evaluate this by comparing mock behavior against the original implementation and classify each mock into one of three categories:
\emph{over-approximating (OB)}, where the mock exhibits behaviors not present in the original implementation;
\emph{under-approximating (UB)}, where it omits valid behaviors; and
\emph{correct behavior (CB)}, where it closely matches the original protocol.

We compute both harness accuracy and mock accuracy metrics via manual analysis.
To ensure scalability, we perform random sampling by selecting one protocol from each subsystem category (five in total)\extonly{ and manually evaluate the corresponding harnesses and mocks}.

\subsubsection{Results.}
\label{subsubsec:effectresults}
\tbl{tab:metistable} reports the evaluation metrics across five protocols\extonly{ sampled from different subsystem categories}.
The harness-generation metrics (first set of columns) achieve perfect scores, indicating that \systemname{} successfully generates harnesses capable of exercising the interface functions.

The quality of generated mocks, however, varies.
Out of 220 generated mock methods, 39 correctly mimic the original behavior (CB), 28 only under-approximate it (UB), 19 only over-approximate it (OB), and the remaining 134 do both (OB\&UB).
Over-approximation most commonly comes from missing validation and state tracking: a mock returns success for arguments or protocol states that the real implementation rejects, or returns outputs that violate invariants the real producer guarantees.
Manual inspection revealed that most UB cases arise from asynchronous callback handling.
Although \systemname{} provides the necessary \inlinecode{AsyncCallback} APIs, current \acp{LLM} struggle to fully model the associated asynchronous semantics.


\begin{wrapfigure}{R}{0.45\textwidth}
\vspace{-\baselineskip}
\centering
\begin{tikzpicture}
\begin{axis}[
  width=1.03\linewidth, height=4.6cm,
  xmin=0, xmax=5, ymin=0, ymax=100,
  xtick={0,1,2,3,4,5},
  ytick={0,20,40,60,80,100},
  xticklabel style={font=\scriptsize},
  yticklabel style={font=\scriptsize},
  xlabel={Refinement round},
  ylabel={Line-coverage gain (\%)},
  xlabel style={font=\footnotesize}, ylabel style={font=\footnotesize},
  ymajorgrids, grid style={dashed, gray!40},
  legend style={font=\scriptsize, at={(0.5,1.03)}, anchor=south, legend columns=3, draw=none,
                /tikz/every even column/.append style={column sep=6pt}},
  legend cell align=left,
]
\addplot[gray!45, very thin, forget plot] coordinates {(0,0.0) (1,2.6) (2,9.6) (3,9.6) (4,9.6) (5,9.6)};
\addplot[gray!45, very thin, forget plot] coordinates {(0,0.0) (1,74.6) (2,74.6) (3,74.6) (4,74.6) (5,74.6)};
\addplot[gray!45, very thin, forget plot] coordinates {(0,0.0) (1,47.9) (2,61.9) (3,61.9) (4,61.9) (5,61.9)};
\addplot[gray!45, very thin, forget plot] coordinates {(0,0.0) (1,0.0) (2,0.0) (3,0.0) (4,0.0) (5,0.0)};
\addplot[gray!45, very thin, forget plot] coordinates {(0,0.0) (1,0.1) (2,0.1) (3,0.1) (4,0.1) (5,0.1)};
\addplot[gray!45, very thin, forget plot] coordinates {(0,0.0) (1,0.0) (2,0.0) (3,0.0) (4,0.0) (5,0.0)};
\addplot[gray!45, very thin, forget plot] coordinates {(0,0.0) (1,37.4) (2,37.4) (3,37.4) (4,37.4) (5,37.4)};
\addplot[gray!45, very thin, forget plot] coordinates {(0,0.0) (1,59.0) (2,59.0) (3,59.0) (4,59.0) (5,59.0)};
\addplot[gray!45, very thin, forget plot] coordinates {(0,0.0) (1,52.5) (2,71.8) (3,72.9) (4,79.1) (5,79.1)};
\addplot[gray!45, very thin, forget plot] coordinates {(0,0.0) (1,1.0) (2,1.0) (3,1.0) (4,1.0) (5,1.0)};
\addplot[gray!45, very thin, forget plot] coordinates {(0,0.0) (1,78.9) (2,78.9) (3,78.9) (4,78.9) (5,78.9)};
\addplot[gray!45, very thin, forget plot] coordinates {(0,0.0) (1,19.0) (2,38.4) (3,38.4) (4,38.4) (5,38.4)};
\addplot[gray!45, very thin, forget plot] coordinates {(0,0.0) (1,0.0) (2,0.0) (3,0.0) (4,0.0) (5,0.0)};
\addplot[gray!45, very thin, forget plot] coordinates {(0,0.0) (1,74.6) (2,74.6) (3,74.6) (4,74.6) (5,74.6)};
\addplot[gray!45, very thin, forget plot] coordinates {(0,0.0) (1,65.4) (2,65.4) (3,65.4) (4,65.4) (5,65.4)};
\addplot[gray!45, very thin, forget plot] coordinates {(0,0.0) (1,46.5) (2,74.3) (3,74.3) (4,74.3) (5,74.3)};
\addplot[gray!45, very thin, forget plot] coordinates {(0,0.0) (1,42.2) (2,42.2) (3,42.2) (4,42.2) (5,42.2)};
\addplot[gray!45, very thin, forget plot] coordinates {(0,0.0) (1,53.2) (2,53.2) (3,53.2) (4,53.2) (5,53.2)};
\addplot[gray!45, very thin, forget plot] coordinates {(0,0.0) (1,0.8) (2,4.6) (3,5.9) (4,5.9) (5,5.9)};
\addplot[gray!45, very thin, forget plot] coordinates {(0,0.0) (1,0.0) (2,0.0) (3,0.0) (4,0.0) (5,0.0)};
\addplot[gray!45, very thin, forget plot] coordinates {(0,0.0) (1,0.0) (2,0.0) (3,0.0) (4,0.0) (5,0.0)};
\addplot[gray!45, very thin, forget plot] coordinates {(0,0.0) (1,74.4) (2,93.9) (3,93.9) (4,93.9) (5,93.9)};
\addplot[gray!45, very thin, forget plot] coordinates {(0,0.0) (1,0.0) (2,0.0) (3,0.0) (4,0.0) (5,0.0)};
\addplot[gray!45, very thin, forget plot] coordinates {(0,0.0) (1,69.9) (2,73.7) (3,73.7) (4,73.7) (5,73.7)};
\addplot[gray!45, very thin, forget plot] coordinates {(0,0.0) (1,11.0) (2,23.7) (3,23.7) (4,23.7) (5,23.7)};
\addplot[gray!45, very thin, forget plot] coordinates {(0,0.0) (1,37.5) (2,37.5) (3,37.5) (4,37.5) (5,37.5)};
\addplot[gray!45, very thin, forget plot] coordinates {(0,0.0) (1,0.5) (2,11.4) (3,22.0) (4,24.1) (5,25.9)};
\addplot[gray!45, very thin, forget plot] coordinates {(0,0.0) (1,4.2) (2,20.3) (3,33.4) (4,33.4) (5,33.4)};
\addplot[gray!45, very thin, forget plot] coordinates {(0,0.0) (1,96.3) (2,96.3) (3,96.3) (4,96.3) (5,96.3)};
\addplot[gray!45, very thin, forget plot] coordinates {(0,0.0) (1,0.3) (2,1.8) (3,16.3) (4,17.3) (5,19.3)};
\addplot[gray!45, very thin, forget plot] coordinates {(0,0.0) (1,7.1) (2,7.1) (3,7.1) (4,7.1) (5,7.1)};
\addplot[gray!45, very thin, forget plot] coordinates {(0,0.0) (1,63.0) (2,74.5) (3,74.5) (4,74.5) (5,74.5)};
\addplot[gray!45, very thin, forget plot] coordinates {(0,0.0) (1,5.7) (2,10.0) (3,10.0) (4,10.0) (5,10.0)};
\addplot[gray!45, very thin, forget plot] coordinates {(0,0.0) (1,1.0) (2,7.9) (3,7.9) (4,7.9) (5,7.9)};
\addplot[gray!45, very thin, forget plot] coordinates {(0,0.0) (1,3.2) (2,3.2) (3,3.2) (4,3.2) (5,3.2)};
\addplot[gray!45, very thin, forget plot] coordinates {(0,0.0) (1,11.8) (2,35.9) (3,38.4) (4,40.2) (5,40.8)};
\addplot[gray!45, very thin, forget plot] coordinates {(0,0.0) (1,6.4) (2,6.4) (3,6.4) (4,6.4) (5,6.4)};
\addplot[gray!45, very thin, forget plot] coordinates {(0,0.0) (1,9.1) (2,20.6) (3,28.7) (4,28.7) (5,28.7)};
\addplot[gray!45, very thin, forget plot] coordinates {(0,0.0) (1,22.4) (2,22.4) (3,22.4) (4,22.4) (5,22.4)};
\addplot[gray!45, very thin, forget plot] coordinates {(0,0.0) (1,12.2) (2,16.7) (3,19.9) (4,19.9) (5,19.9)};
\addplot[gray!45, very thin, forget plot] coordinates {(0,0.0) (1,0.0) (2,0.0) (3,0.0) (4,0.0) (5,0.0)};
\addplot[gray!45, very thin, forget plot] coordinates {(0,0.0) (1,48.8) (2,48.8) (3,48.8) (4,48.8) (5,48.8)};
\addplot[gray!45, very thin, forget plot] coordinates {(0,0.0) (1,15.9) (2,27.7) (3,27.7) (4,27.7) (5,27.7)};
\addplot[gray!45, very thin, forget plot] coordinates {(0,0.0) (1,0.0) (2,0.0) (3,0.0) (4,0.0) (5,0.0)};
\addplot[gray!45, very thin, forget plot] coordinates {(0,0.0) (1,36.1) (2,36.1) (3,36.1) (4,36.1) (5,36.1)};
\addplot[gray!45, very thin, forget plot] coordinates {(0,0.0) (1,1.6) (2,12.1) (3,12.1) (4,12.1) (5,12.1)};
\addplot[gray!45, very thin, forget plot] coordinates {(0,0.0) (1,9.9) (2,14.2) (3,15.8) (4,19.3) (5,20.9)};
\addplot[gray!45, very thin, forget plot] coordinates {(0,0.0) (1,0.0) (2,0.0) (3,0.0) (4,0.0) (5,0.0)};
\addplot[gray!45, very thin, forget plot] coordinates {(0,0.0) (1,0.0) (2,0.0) (3,0.0) (4,0.0) (5,0.0)};
\addplot[gray!45, very thin, forget plot] coordinates {(0,0.0) (1,4.3) (2,39.7) (3,40.5) (4,40.5) (5,40.5)};
\addplot[gray!45, very thin, forget plot] coordinates {(0,0.0) (1,33.7) (2,51.4) (3,51.4) (4,51.4) (5,51.4)};
\addplot[gray!45, very thin, forget plot] coordinates {(0,0.0) (1,22.1) (2,22.1) (3,22.1) (4,22.1) (5,22.1)};
\addplot[gray!45, very thin, forget plot] coordinates {(0,0.0) (1,9.9) (2,9.9) (3,9.9) (4,9.9) (5,9.9)};
\addplot[gray!45, very thin, forget plot] coordinates {(0,0.0) (1,0.0) (2,0.0) (3,0.0) (4,0.0) (5,0.0)};
\addplot[gray!45, very thin, forget plot] coordinates {(0,0.0) (1,28.6) (2,33.6) (3,33.6) (4,33.6) (5,33.6)};
\addplot[gray!45, very thin, forget plot] coordinates {(0,0.0) (1,82.8) (2,84.8) (3,84.8) (4,84.8) (5,84.8)};
\addplot[gray!45, very thin, forget plot] coordinates {(0,0.0) (1,16.5) (2,30.9) (3,30.9) (4,30.9) (5,30.9)};
\addplot[gray!45, very thin, forget plot] coordinates {(0,0.0) (1,0.0) (2,0.0) (3,0.0) (4,0.0) (5,0.0)};
\addplot[gray!45, very thin, forget plot] coordinates {(0,0.0) (1,0.0) (2,0.0) (3,0.0) (4,0.0) (5,0.0)};
\addlegendimage{gray!60, thin}
\addlegendentry{Driver}
\addplot[thick, blue!70!black, mark=*, mark size=1.5pt] coordinates {(0,0.0) (1,23.8) (2,29.2) (3,30.1) (4,30.4) (5,30.5)};
\addlegendentry{Mean}
\addplot[thick, dashed, black, mark=square*, mark size=1.3pt] coordinates {(0,0.0) (1,9.9) (2,20.6) (3,22.4) (4,23.7) (5,23.7)};
\addlegendentry{Median}
\end{axis}
\end{tikzpicture}
\vspace{-0.5\baselineskip}
\caption{Cumulative increase in driver line coverage after each refinement round.}
\label{fig:refinement}
\vspace{-\baselineskip}
\end{wrapfigure}
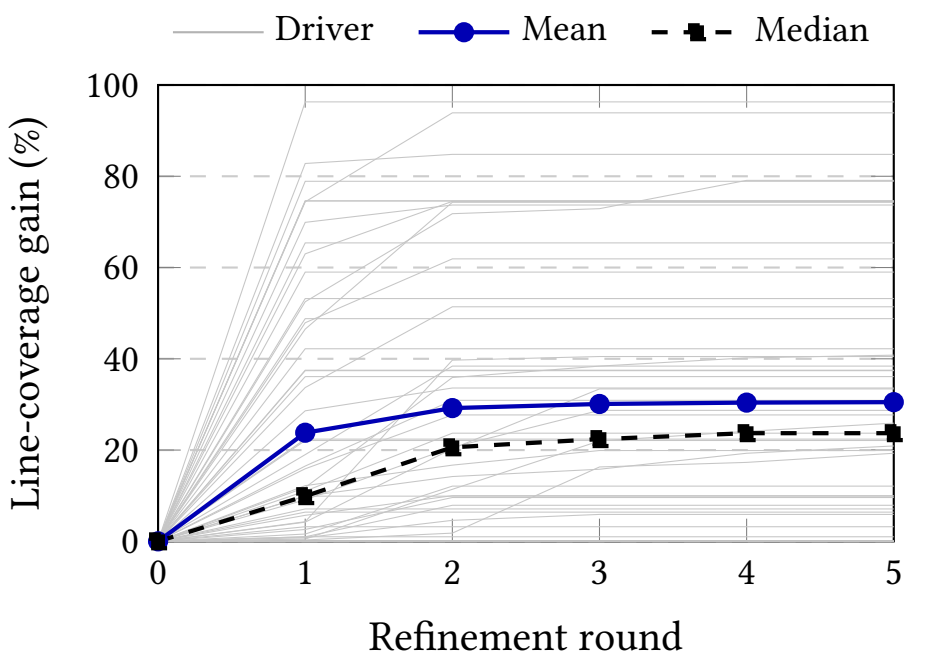

\parahead{Refinement}
\fig{fig:refinement} shows the effect of the refinement loop on the 59 drivers that built in our campaign with Claude Sonnet~5.5.
Each round fuzzes the current harness and mocks for 120\,s, measures driver line coverage, and asks the \ac{LLM} to revise the harness or a mock then a revision is kept only if it covers at least four new driver lines.
In round~1, 46 of the 59 drivers kept a revision (30 of them revised the harness and 17 a mock), adding 30.5\% of line coverage on average.
Rounds 2-5 kept 26, 10, 5, and 4 revisions, adding 12.3, 5.7, 2.9, and 1.5\% on average.
Overall, refinement raised the mean driver line coverage at the 120\,s probe from 33.4\% to 63.9\% (median 28.6\% to 73.3\%) where 46 of the 59 drivers improved, and most of the gain arrives in the first two rounds.

\begin{table}[t]
\centering
\scriptsize
\setlength{\tabcolsep}{2pt}
\renewcommand{\arraystretch}{1.05}
\caption{Quality of harnesses and mocks generated by \systemname{} for
one representative driver per subsystem category.
Results discussed in \sect{subsubsec:evalrehostingmethod}.}
\label{tab:metistable}
\resizebox{\columnwidth}{!}{%
\begin{tabular}{|l|r|cccc|r|rrr|}
\hline
\textbf{Driver}
  & \textbf{Num(Fp)}
  & \textbf{Com} & \textbf{Arg} & \textbf{DepH} & \textbf{CArg}
  & \textbf{\begin{tabular}[c]{@{}c@{}}Total Dep.\\ Funcs\end{tabular}}
  & \textbf{OB} & \textbf{UB} & \textbf{CB} \\
\hline
\texttt{Dhcp4Dxe}            &  9 & 9/9 (100\%)   & 9/9 (100\%)   & 3/3 (100\%) & 9/9 (100\%)     &  84 & 10 &  9 & 16 \\ \hline
\texttt{PartitionDxe}        &  4 & 4/4 (100\%)   & 4/4 (100\%)   & 0/0 (N/A)   & 4/4 (100\%)     &  37 &  2 &  5 &   8 \\ \hline
\texttt{UsbBusDxe}           & 13 & 13/13 (100\%) & 13/13 (100\%) & 4/4 (100\%) & 15/16 (93.8\%)  &  44 &  4 &  6 & 10 \\ \hline
\texttt{GraphicsConsoleDxe}  &  9 & 9/9 (100\%)   & 9/9 (100\%)   & 0/0 (N/A)   & 2/2 (100\%)     &  51 &  3 &  8 &  5 \\ \hline
\texttt{SmmControl2Dxe}      &  2 & 2/2 (100\%)   & 2/2 (100\%)   & 0/0 (N/A)   & 0/0 (N/A)       &   4 &  0 &  0 &   0 \\ \hline
\textbf{Mean / Total}        & \textbf{7.4} & \textbf{37/37 (100\%)} & \textbf{37/37 (100\%)} & \textbf{7/7 (100\%)} & \textbf{30/31 (96.8\%)} & \textbf{220} & \textbf{19} & \textbf{28} & \textbf{39} \\ \hline
\end{tabular}%
}
\end{table}

\subsection{RQ2: Fuzzing Effectiveness}
\label{sec:eval-rq2}

\extonly{We want to assess the effectiveness of \systemname{} pipeline in testing \edk{} protocols.}
\subsubsection{Method}
We run the full \systemname{} pipeline on all \numdrivers drivers and evaluate its effectiveness in terms of code coverage, rediscovery of previously known bugs, and identification of new vulnerabilities.
We report both line and function coverage of the target driver's own sources.
Since prior work (\eg{} \fuzzuer{}, \hbfa{}) reports only line coverage, we compare against them on line coverage.

To evaluate known bug detection, we check whether \systemname{} triggers the corresponding crash with the correct root cause.
For newly discovered bugs, we report a tuple of the form \texttt{uniq}/\texttt{distinct}/\texttt{bugs}, representing:
(i) the number of unique crashes reported by \afl{},
(ii) the number of distinct crashes, \ie{} crashes that reproduce, grouped by fingerprint, and
(iii) the number of driver bugs, \ie{} distinct crashes attributed to the target driver rather than to a mock or the harness.

\extonly{For example, a result of \texttt{224/9/4} indicates that \afl{} reported 224 unique crashes, which reduced to 9 distinct crashes, 4 of which are bugs in the driver itself.}

\extonly{As mentioned before (\sect{subsubsec:llmchoices}), we report our results by using the Claude Sonnet~5.5 LLM model.}




\subsubsection{Results}
\begin{table*}[]
\centering
\tiny
\setlength{\tabcolsep}{1pt}
\renewcommand{\arraystretch}{0.65}
\caption{Per-driver fuzzing results: \systemname{} (LF mode) vs the LLM-only baseline (LLM), \fuzzuer{}~\cite{glosner_fuzzuer_2025}, and the hand-written \hbfa{}~\cite{hbfa} harness where one exists. Coverage is measured on the target driver's own sources; L = line coverage, F = function coverage (\fuzzuer{} and \hbfa{} report line coverage only). ``*'' = failed to generate working harness + mock. ``---'' = data not available. New Findings for \systemname{} and LLM = AFL unique crashes / distinct crashes (reproduced, grouped by fingerprint) / driver bugs (distinct crashes attributed to the driver, not a mock or the harness). Mean (measured) averages over the drivers with measured coverage (\systemname{}: 59, LLM: \numllmmeasured{}); Mean (all \numdrivers{}) averages over all \numdrivers{} drivers, counting failures (``*'') as 0. Column semantics are discussed in \sect{sec:eval-rq2} and \sect{sec:eval-rq3}.}
\label{tab:fuzzingeval}
\begin{tabular}{l | rrr | r@{\hspace{3pt}}r@{\hspace{6pt}}r@{\hspace{3pt}}r@{\hspace{6pt}}c@{\hspace{4pt}}c | r c cc | r c cc}
\toprule
 & \multicolumn{3}{c|}{\textbf{Iface Fns}}
 & \multicolumn{6}{c|}{\textbf{Coverage (\%)}}
 & \multicolumn{4}{c|}{\textbf{Prev. Known Bugs}}
 & \multicolumn{4}{c}{\textbf{New Findings (uniq/distinct/bugs)}} \\[1pt]
 & & &
 & \multicolumn{2}{c}{\systemname{}} & \multicolumn{2}{c}{LLM} & {\fuzzuer{}} & \multicolumn{1}{c|}{\hbfa{}}
 & & & & & & & & \\[1pt]
\cmidrule(lr){5-6}\cmidrule(lr){7-8}
\textbf{Driver}
 & $\mid\protocolfunctionset{\targetprotcol{}}\mid$ & $ComF$ & $VoidF$
 & L & F & L & F & L & L
 & {\systemname{}} & {LLM} & {\fuzzuer{}} & {\hbfa{}}
 & {\systemname{}} & {LLM} & {\fuzzuer{}} & {\hbfa{}} \\
\midrule
\multicolumn{18}{l}{\emph{NetworkPkg}} \\
\midrule
\texttt{ArpDxe} & 9 & 9 & 5 & 89.6 & 91.7 & * & * & 20.9 & --- & $\times$ & * & $\times$ & --- & 0/0/0 & * & 0 & --- \\
\texttt{Dhcp4Dxe} & 11 & 11 & 0 & 89.1 & 95.3 & * & * & 37.3 & --- & --- & --- & --- & --- & 224/9/4 & * & 0 & --- \\
\texttt{Dhcp6Dxe} & 11 & 11 & 1 & 73.0 & 83.8 & * & * & 31.8 & --- & \checkmark & * & $\times$ & --- & 187/5/4 & * & 0 & --- \\
\texttt{DnsDxe} & 20 & 20 & 0 & 79.3 & 94.6 & * & * & 9.5 & --- & \checkmark & * & $\times$ & --- & 316/5/4 & * & 0 & --- \\
\texttt{DpcDxe} & 2 & 2 & 1 & 94.4 & 100.0 & 100.0 & 100.0 & 87.7 & --- & --- & --- & --- & --- & 0/0/0 & 34/1/1 & 0 & --- \\
\texttt{HttpBootDxe} & 5 & 5 & 2 & 71.0 & 87.0 & * & * & 29.7 & --- & --- & --- & --- & --- & 144/7/6 & * & 0 & --- \\
\texttt{HttpDxe} & 8 & 8 & 0 & 78.9 & 94.7 & * & * & 5.7 & --- & $\times$ & * & $\times$ & --- & 309/10/7 & * & 0 & --- \\
\texttt{IScsiDxe} & 17 & 17 & 4 & 14.8 & 26.7 & * & * & 13.0 & --- & --- & --- & --- & --- & 8/1/0 & * & 0 & --- \\
\texttt{Ip4Dxe} & 17 & 17 & 2 & 84.3 & 97.0 & * & * & 36.9 & --- & \checkmark & * & $\times$ & --- & 216/9/4 & * & 0 & --- \\
\texttt{Ip6Dxe} & 15 & 15 & 2 & * & * & * & * & 35.3 & --- & * & * & $\times$ & --- & * & * & 0 & --- \\
\texttt{MnpDxe} & 13 & 13 & 0 & 82.8 & 98.6 & * & * & 58.3 & --- & --- & --- & --- & --- & 109/1/1 & * & 0 & --- \\
\texttt{Mtftp4Dxe} & 10 & 10 & 0 & 89.2 & 98.4 & * & * & 26.4 & --- & --- & --- & --- & --- & 172/8/6 & * & 0 & --- \\
\texttt{Mtftp6Dxe} & 10 & 10 & 0 & 74.3 & 85.7 & * & * & 15.3 & --- & --- & --- & --- & --- & 69/6/2 & * & 0 & --- \\
\texttt{SnpDxe} & 13 & 13 & 4 & 18.8 & 19.1 & * & * & 70.0 & --- & --- & --- & --- & --- & 0/0/0 & * & 0 & --- \\
\texttt{TcpDxe} & 30 & 30 & 0 & 55.2 & 66.5 & * & * & 11.9 & --- & $\times$ & * & $\times$ & --- & 96/5/5 & * & 0 & --- \\
\texttt{TlsAuthConfigDxe} & 3 & 3 & 0 & 62.4 & 61.9 & 88.0 & 100.0 & 22.3 & --- & --- & --- & --- & --- & 1/1/1 & 0/0/0 & 0 & --- \\
\texttt{TlsDxe} & 8 & 8 & 4 & 85.4 & 87.5 & * & * & 37.0 & --- & --- & --- & --- & --- & 94/3/3 & * & 0 & --- \\
\texttt{Udp4Dxe} & 10 & 10 & 0 & 82.3 & 87.5 & * & * & 50.9 & --- & --- & --- & --- & --- & 0/0/0 & * & 0 & --- \\
\texttt{Udp6Dxe} & 9 & 9 & 0 & 80.8 & 87.5 & * & * & 47.8 & --- & --- & --- & --- & --- & 124/4/3 & * & 0 & --- \\
\texttt{UefiPxeBcDxe} & 14 & 14 & 4 & 67.8 & 79.8 & * & * & 28.5 & --- & \checkmark & * & $\times$ & --- & 312/6/4 & * & 0 & --- \\
\texttt{VlanConfigDxe} & 3 & 3 & 0 & 93.9 & 100.0 & 95.1 & 100.0 & 45.1 & --- & --- & --- & --- & --- & 0/0/0 & 0/0/0 & 0 & --- \\
\texttt{WifiConnectionMgr} & 3 & 3 & 0 & 75.9 & 81.4 & * & * & 7.8 & --- & --- & --- & --- & --- & 209/4/1 & * & 0 & --- \\
\midrule
\multicolumn{18}{l}{\emph{PCI/Storage}} \\
\midrule
\texttt{AtaAtapiPassThru} & 14 & 14 & 0 & 70.8 & 84.7 & * & * & 43.7 & --- & --- & --- & --- & --- & 62/2/1 & * & 0 & --- \\
\texttt{AtaBusDxe} & 14 & 14 & 9 & 83.9 & 94.7 & * & * & 42.4 & --- & --- & --- & --- & --- & 0/0/0 & * & 0 & --- \\
\texttt{DiskIoDxe} & 6 & 6 & 4 & 86.6 & 100.0 & 87.7 & 94.7 & 55.1 & --- & --- & --- & --- & --- & 79/5/0 & 77/3/3 & 0 & --- \\
\texttt{EmmcDxe} & 15 & 15 & 9 & 88.7 & 95.3 & 85.7 & 100.0 & 6.3 & --- & --- & --- & --- & --- & 94/1/1 & 116/5/2 & 1 & --- \\
\texttt{IncompatPciDevSup} & 1 & 1 & 1 & 81.0 & 100.0 & 81.0 & 100.0 & 83.3 & --- & --- & --- & --- & --- & 0/0/0 & 0/0/0 & 0 & --- \\
\texttt{NvmExpressDxe} & 21 & 21 & 10 & 85.5 & 97.0 & * & * & --- & --- & --- & --- & --- & --- & 124/7/7 & * & --- & --- \\
\texttt{PartitionDxe} & 8 & 8 & 4 & 62.6 & 92.1 & * & * & 53.9 & 15.2 & $\times$ & * & $\times$ & $\times$ & 30/1/0 & * & 0 & 0 \\
\texttt{PciBusDxe} & 15 & 15 & 6 & 71.1 & 85.5 & * & * & 47.7 & --- & --- & --- & --- & --- & 295/5/4 & * & 1 & --- \\
\texttt{PciHostBridgeDxe} & 19 & 19 & 9 & 50.4 & 75.0 & * & * & 84.4 & --- & --- & --- & --- & --- & 0/0/0 & * & 0 & --- \\
\texttt{RamDiskDxe} & 13 & 13 & 4 & 52.9 & 71.4 & 61.9 & 75.0 & 17.0 & --- & --- & --- & --- & --- & 43/2/1 & 121/1/1 & 0 & --- \\
\texttt{ScsiBusDxe} & 5 & 5 & 0 & 83.4 & 100.0 & 0.0 & 0.0 & 72.6 & --- & --- & --- & --- & --- & 39/1/0 & 88/1/1 & 0 & --- \\
\texttt{ScsiDiskDxe} & 15 & 15 & 9 & 80.6 & 92.2 & 82.6 & 95.3 & 21.9 & --- & --- & --- & --- & --- & 301/7/7 & 272/2/1 & 0 & --- \\
\texttt{SdDxe} & 13 & 13 & 7 & 90.1 & 100.0 & * & * & 9.1 & --- & --- & --- & --- & --- & 40/4/4 & * & 1 & --- \\
\texttt{SdMmcPciHcDxe} & 5 & 5 & 0 & 83.9 & 97.1 & * & * & 14.6 & --- & $\times$ & * & $\times$ & --- & 0/0/0 & * & 0 & --- \\
\texttt{SataControllerDxe} & 6 & 6 & 0 & 89.7 & 100.0 & 99.6 & 100.0 & 71.4 & --- & --- & --- & --- & --- & 56/7/0 & 0/0/0 & 0 & --- \\
\texttt{UfsPassThruDxe} & 10 & 10 & 0 & 81.2 & 96.5 & * & * & 5.9 & --- & --- & --- & --- & --- & 0/0/0 & * & 0 & --- \\
\midrule
\multicolumn{18}{l}{\emph{USB}} \\
\midrule
\texttt{EhciDxe} & 13 & 13 & 6 & 83.9 & 89.9 & * & * & 67.3 & --- & --- & --- & --- & --- & 110/2/2 & * & 0 & --- \\
\texttt{UhciDxe} & 13 & 13 & 6 & 72.3 & 90.4 & 91.4 & 98.6 & 74.2 & --- & --- & --- & --- & --- & 0/0/0 & 0/0/0 & 0 & --- \\
\texttt{UsbBusDxe} & 13 & 13 & 6 & 76.6 & 93.9 & * & * & 60.2 & --- & --- & --- & --- & --- & 131/14/4 & * & 0 & --- \\
\texttt{UsbKbDxe} & 7 & 7 & 2 & 94.9 & 100.0 & 96.7 & 100.0 & 49.0 & --- & --- & --- & --- & --- & 0/0/0 & 82/1/1 & 0 & --- \\
\texttt{UsbMassStorageDxe} & 8 & 8 & 5 & 93.1 & 100.0 & 61.5 & 74.5 & 49.6 & --- & --- & --- & --- & --- & 94/1/1 & 61/1/1 & 0 & --- \\
\texttt{UsbMouseDxe} & 2 & 2 & 0 & 93.4 & 100.0 & 96.6 & 100.0 & 64.9 & --- & --- & --- & --- & --- & 51/1/0 & 16/3/3 & 0 & --- \\
\texttt{XhciDxe} & 13 & 13 & 6 & 83.2 & 93.0 & * & * & 5.6 & --- & --- & --- & --- & --- & 149/11/11 & * & 1 & --- \\
\midrule
\multicolumn{18}{l}{\emph{Console/HII}} \\
\midrule
\texttt{ConSplitterDxe} & 23 & 23 & 2 & 76.0 & 100.0 & 90.9 & 100.0 & 59.3 & --- & --- & --- & --- & --- & 0/0/0 & 262/13/13 & 0 & --- \\
\texttt{DisplayEngineDxe} & 4 & 2 & 0 & 9.6 & 19.3 & * & * & 4.7 & --- & --- & --- & --- & --- & 9/1/1 & * & 0 & --- \\
\texttt{EnglishDxe} & 6 & 6 & 0 & 100.0 & 100.0 & 100.0 & 100.0 & 76.8 & --- & --- & --- & --- & --- & 40/1/1 & 68/1/1 & 0 & --- \\
\texttt{GraphicsConsoleDxe} & 9 & 9 & 0 & 89.2 & 100.0 & 0.0 & 0.0 & 74.8 & --- & --- & --- & --- & --- & 0/0/0 & 0/0/0 & 0 & --- \\
\texttt{GraphicsOutputDxe} & 3 & 3 & 0 & 81.1 & 100.0 & 96.7 & 100.0 & --- & --- & --- & --- & --- & --- & 63/3/0 & 95/4/4 & --- & --- \\
\texttt{HiiDatabaseDxe} & 39 & 39 & 0 & 46.8 & 70.7 & * & * & 38.6 & --- & --- & --- & --- & --- & 311/31/29 & * & 3 & --- \\
\texttt{HiiResourcesSample} & 0 & 0 & 0 & 42.1 & 100.0 & 0.0 & 0.0 & --- & --- & --- & --- & --- & --- & 0/0/0 & 0/0/0 & --- & --- \\
\texttt{LogoDxe} & 1 & 1 & 0 & 100.0 & 100.0 & 100.0 & 100.0 & 83.0 & --- & --- & --- & --- & --- & 0/0/0 & 0/0/0 & 0 & --- \\
\texttt{SetupBrowserDxe} & 12 & 4 & 0 & 35.2 & 62.5 & * & * & 3.0 & --- & --- & --- & --- & --- & 160/3/0 & * & 1 & --- \\
\midrule
\multicolumn{18}{l}{\emph{SMM/Firmware}} \\
\midrule
\texttt{FaultTolerantWrite} & 6 & 6 & 2 & 87.9 & 100.0 & * & * & --- & --- & --- & --- & --- & --- & 100/6/6 & * & --- & --- \\
\texttt{S3SaveStateDxe} & 4 & 4 & 0 & 95.2 & 100.0 & 96.7 & 100.0 & 69.8 & --- & --- & --- & --- & --- & 75/1/1 & 111/1/0 & 0 & --- \\
\texttt{SmmControl2Dxe} & 2 & 2 & 0 & 37.0 & 71.4 & 71.5 & 100.0 & 77.8 & --- & --- & --- & --- & --- & 0/0/0 & 0/0/0 & 0 & --- \\
\texttt{SmmS3SaveState} & 4 & 4 & 0 & 97.7 & 100.0 & 99.7 & 100.0 & --- & --- & --- & --- & --- & --- & 70/3/1 & 0/0/0 & --- & --- \\
\midrule
\multicolumn{18}{l}{\emph{Utility}} \\
\midrule
\texttt{PrintDxe} & 10 & 0 & 0 & 28.6 & 33.3 & 100.0 & 100.0 & 75.0 & --- & --- & --- & --- & --- & 0/0/0 & 67/11/0 & 0 & --- \\
\texttt{RestJsonStructure} & 4 & 4 & 0 & 2.1 & 11.1 & 97.9 & 100.0 & 15.8 & --- & --- & --- & --- & --- & 0/0/0 & 0/0/0 & 0 & --- \\
\midrule
\textbf{Mean (measured) / Total} & \textbf{617} & \textbf{597} & \textbf{136} & \textbf{73.1} & \textbf{85.5} & \textbf{79.2} & \textbf{85.5} & \textbf{42.0} & \textbf{15.2} & \textbf{4} & \textbf{0} & \textbf{0} & \textbf{0} & \textbf{5116/204/137} & \textbf{1470/48/32} & \textbf{8} & \textbf{0} \\
\textbf{Mean (all \numdrivers{})} & & & & \textbf{71.9} & \textbf{84.0} & \textbf{33.0} & \textbf{35.6} & & & & & & & & & & \\
\bottomrule
\end{tabular}
\end{table*}

\tbl{tab:fuzzingeval} shows the results across all protocols.

\parahead{Coverage}
The average line and function coverage across the 59 drivers with measured coverage is \avgcodecvgperc{} and \avgfncvgperc{}, respectively.
The remaining coverage gap primarily stems from under-approximation in dependency mocks, particularly for asynchronous interactions.
We observe no strong correlation between interface-function complexity and achieved coverage.
For example, {\tt EmmcDxe}, which exposes 15 complex interface functions, achieves 88.7\% line coverage, whereas {\tt HttpBootDxe}, with only 5 complex functions, achieves 71.0\% line coverage.
The primary difference lies in the complexity of their dependencies: {\tt HttpBootDxe} depends on substantially more complex protocols than {\tt EmmcDxe}.\extonly{
Although our generated harnesses successfully exercise all interface functions, under-approximated dependency behavior limits deeper exploration of protocol logic and ultimately reduces coverage.}

\parahead{Previously known bugs and False Negatives}
\extonly{As shown in the \emph{Previously Known Bugs} column, }\systemname{} rediscovered \numrediscoveredbugs{} of the \numpreviousbugs{} previously known vulnerabilities in the evaluated drivers, a false-negative rate of \percfnrate{}.
Three of them crash at the exact vulnerable site (CVE-2018-12178 in \texttt{DnsDxe}, CVE-2019-14559 in \texttt{Ip4Dxe}, and CVE-2023-45234 in \texttt{UefiPxeBcDxe}); the fourth (CVE-2023-45229 in \texttt{Dhcp6Dxe}) crashes on the same Advertise-processing path rather than at the vulnerable check.
\extonly{\Apdxref{\tbl{tab:new_bugs}} lists each known bug with its crash site or, if missed, the reason.}

The ten misses fall into three groups.
First, four bugs were triggered but are invisible to our crash-based oracle (AddressSanitizer and assertions): predictable TCP initial sequence numbers (CVE-2023-45236), missing TLS host-name verification (CVE-2019-14553), a receive-buffer leak (CVE-2019-14559 in \texttt{ArpDxe}), and a double \texttt{Unmap} that the mock silently rejects (CVE-2019-14587).
Detecting these requires semantic oracles rather than memory-safety checks.
Second, the three \texttt{Ip6Dxe} bugs were never fuzzed: because \afl{} requires at least one non-crashing seed input, we deterministically check for one before fuzzing, and since the \texttt{Ip6Dxe} harness crashed during startup, before reaching the driver's entry point, no input passed this check.
Two of these bugs are also infinite loops, which would additionally require hang detection.
Third, three bugs need inputs that our harnesses and mocks did not produce: CVE-2018-12180 in \texttt{PartitionDxe} needs a block size below 512 bytes, which the harness never generates; CVE-2023-45230 needs an oversized DHCPv6 Server-ID, whereas the mock injects a fixed 8-byte one; and CVE-2023-45235 needs a DHCPv6 proxy offer that fuzzing never generated.

\parahead{New bugs and False Positives}\extonly{
The last columns of \tbl{tab:fuzzingeval} report the crash and bug statistics.}
Overall, \afl{} reported \numafluniquecrashes{} unique crashes, which reduced to \numdistinctcrashes{} distinct crashes, of which \numdriverbugs{} are attributed to the target drivers.

\parahead{Responsible Disclosure}
We are in the process of reporting these \numdriverbugs{} bugs\extonly{, which takes time because each report requires creating a proof-of-concept (PoC)}.
So far, we have reported \numreportedbugs{} of them, out of which \remotebugs{} are remotely triggerable, and all our reported vulnerabilities have been acknowledged by the vendors.
\extelse{\Apdxref{\tbl{tab:new_bugs}} summarizes the newly discovered vulnerabilities by package; their details are}{The details of the newly discovered vulnerabilities are} withheld pending responsible disclosure and will be published soon.
Many of these are memory-corruption bugs with potentially severe consequences.
Several are in network-facing drivers, such as {\tt Dhcp6Dxe} and {\tt Dhcp4Dxe}, are remotely triggerable.

\subsection{RQ3: Comparative Evaluation}
\label{sec:eval-rq3}
\extonly{Our goal is to compare \systemname{} against state-of-the-art \edk{} fuzzing techniques.}

\subsubsection{Method}
We compare against three representative baselines:
\hbfa{}~\cite{hbfa}, which relies on manually written mocks and harnesses for selected protocols, and
\fuzzuer{}~\cite{glosner_fuzzuer_2025}, a recent technique that fuzzes protocols inside a simulator using automatically generated harnesses.
\ac{LLM}-only baseline, here we used just the \ac{LLM}, without any information about APIs\extonly{ (see \apdxref{Tables~\ref{tab:fuzzcontexlib} and~\ref{tab:asynccallbacklib}})}. We also provided \ac{LLM} with access to all the source code.
For each baseline, we use the authors’ recommended configuration and setup.\extonly{
Similar to \sect{sec:eval-rq2}, we evaluate all approaches using three metrics: code coverage, ability to rediscover known vulnerabilities, and ability to identify new bugs.}

\subsubsection{Results}
\extonly{\tbl{tab:fuzzingeval} summarizes the results across all evaluated protocols.}

\parahead{Comparison with \hbfa{}}
\label{sec:comparison_hbfa}
We use the harnesses available in the latest \hbfa{} commit.
\extonly{As discussed in \sect{sec:background}, \hbfa{} provides only a limited number of harnesses, most of which target individual functions rather than complete protocols.
Consequently, }\hbfa{} was able to exercise only {\tt PartitionDxe} among our evaluated protocols, achieving \hbfalinecvgperc{} line coverage compared to \metispartitionlinecvgperc{} achieved by \systemname{}.
Our findings align with those of Glosner \etal{}~\cite{glosner_fuzzuer_2025}, who also observed that existing \hbfa{} harnesses provide limited coverage.

In terms of bug-finding capability, \hbfa{} neither discovered new bugs nor reproduced the known vulnerability.

\parahead{Comparison with \fuzzuer{}}
\label{sec:comparison_fuzzuer}
\fuzzuer{} executes protocols within a simulator-based environment.
However, not all protocols are available in its simulator configuration; consequently, \fuzzuer{} was able to test only \numfuzzuerprotocols{} out of the \numdrivers{} evaluated drivers.
On the 54 drivers that both tools tested, \systemname{} achieved higher line coverage on 47, with a mean of 72.5\% versus 42.1\% for \fuzzuer{} (median 80.9\% vs. 43.1\%).
\extonly{\fuzzuer{} achieved higher coverage on the remaining seven: \texttt{SnpDxe}, \texttt{IncompatPciDevSup}, \texttt{PciHostBridgeDxe}, \texttt{UhciDxe}, \texttt{SmmControl2Dxe}, \texttt{PrintDxe}, and \texttt{RestJsonStructure}.}

Furthermore, \fuzzuer{} failed to reproduce any known vulnerabilities in these protocols.
Although it reported \numfuzzuerbugs{} new bugs, manual inspection revealed that all were NULL-pointer dereferences.
Many of these crashes are infeasible in practice, as they violate explicit \ac{UEFI} specification requirements that prohibit NULL pointers for the corresponding arguments.
Because \systemname{} incorporates \ac{UEFI} specifications and documentation during harness and mock generation, it avoids generating such invalid inputs and therefore avoids these infeasible crashes.

\parahead{Comparison with \ac{LLM}-baseline}
\label{sec:comparison_llm_baseline}
In the LLM-only experiment, the model's only feedback is \texttt{check\_harness\_build}, which first rejects a harness that breaks basic build rules (e.g., not linking the driver from its own INF), then builds it with the same AFL++/ASan toolchain as the \texttt{build} stage and runs it on an empty input and on each seed, passing it only if the empty-input run and at least one seed run finish without a crash, sanitizer report or hang.
Apart from the raw build and crash output, the model gets no Metis runtime or harness guidance, but because the harness runs as a plain host program with no UEFI environment, each crash points to something the driver is missing, which the model fills in from its own UEFI knowledge and the \texttt{edk2/} sources, so it ends up with a small runtime of its own (e.g., a hand-built \texttt{gBS}).
The resemblance to Metis's runtime is therefore reinvented, and because it is ad hoc and often incomplete, the gap between the two experiments measures what the runtime adds.

What the model rebuilds is far less faithful than Metis's runtime, which ports the EDK-II DXE core's handle database, driver dispatch, events and TPL to the host and generates one fuzz-driven mock per consumed protocol under a fixed contract.

The \ac{LLM} columns in \tbl{tab:fuzzingeval} report the results of the \ac{LLM}-only baseline.
The baseline produced a running harness for only \numllmworking{} of the \numdrivers{} drivers; for the other 35 (``*''), harness generation did not finish for 27, and for 8 the harness failed to build (6) or crashed before fuzzing started (2).
On these \numllmmeasured{} drivers, it reaches \llmlinecvgperc{} line and \llmfncvgperc{} function coverage on average, compared to \metislinecvgonllmset{} and \metisfncvgonllmset{} for \systemname{} on the same drivers; these are largely small drivers with few dependencies.
On three of them (\texttt{ScsiBusDxe}, \texttt{GraphicsConsoleDxe}, and \texttt{HiiResourcesSample}), the baseline covered none of the driver's lines.
Averaged over all \numdrivers{} drivers, counting failures as 0, the baseline reaches 33.0\% line and 35.6\% function coverage, against 71.9\% and 84.0\% for \systemname{}.
The baseline rediscovers none of the \numpreviousbugs{} previously known vulnerabilities, because none of the 10 drivers that contain them got a running harness, and it reports \numllmbugs{} driver bugs, compared to \numdriverbugs{} for \systemname{}.
Of the \numllmbugs{}, manual triage \emph{confirms 16
as true positives (11 distinct bugs). The other 16 are harness artifacts: the
generated harnesses break firmware contracts}.
\textsc{Metis} also finds 3 of these true positives and misses 13
(8 distinct bugs).
These misses have three causes: mocks that are too well-behaved (4 bugs),
no allocation-failure injection (2), and harnesses that never drive the
required event sequences (2).

\extonly{Overall, our results demonstrate that \systemname{} substantially outperforms both \hbfa{} and \fuzzuer{} in terms of coverage and practical bug-finding capability.}

\section{Limitations}
\label{sec:limitations}

\systemname{} has several limitations, both in scope and implementation, that we leave to future work.

\parahead{Bootflow phase coverage}  We target DXE drivers only. 
PEI/SEC, Standalone~MM, and runtime drivers are out of scope.  
SMM-specific vulnerabilities (\eg{} CVE-2017-5753 in \texttt{SmmLockBox}) require a real SMM dispatch context that our user-space rehosting does not emulate.


\parahead{Mock Fidelity} 
\systemname{} currently under-approximates certain dependency behaviors, particularly asynchronous interactions and callback-driven semantics.
As a result, some execution paths and vulnerabilities that depend on precise dependency behavior may not be exercised.
We plan to handle precise modeling of asynchronous callbacks in our future work.

\parahead{Crash Attribution}
Because \systemname{} discards crashes that occur outside the target driver (\eg{} in a mock or the harness), it may miss true bugs whose crash site differs from their root cause, such as a driver passing a malformed buffer that only triggers a crash inside a dependency mock.

\section{Related Work}
\label{sec:relatedwork}

\systemname{} tackles the problem of \edk{} fuzzing by automatically generating the necessary harnesses and mock dependencies. There are prior works that address a similar problem.

\parahead{UEFI/\edk{} Firmware Fuzzing}
Prior work~\cite{mahesh2023validation} attempts to fuzz \ac{UEFI} firmware on real hardware, but such approaches are limited to specific interfaces and do not scale to arbitrary \edk{} components. Scalable \ac{UEFI} firmware fuzzing requires the ability to rehost firmware in an emulated or virtualized environment.
Several works~\cite{simics-fuzzer, uefuzzer, simics-fuzzer-repo, efifuzz, excite, Yin2023rsfuzzer, glosner_fuzzuer_2025} leverage platforms such as Intel Simics~\cite{simics-fuzzer, simics-fuzzer-repo} or Qiling to fuzz \ac{UEFI} firmware.
However, effectively exercising diverse \ac{UEFI}/\edk{} interfaces remains a key challenge.
Consequently, many approaches~\cite{simics-fuzzer, uefuzzer, efifuzz, Yin2023rsfuzzer} focus on post-boot interfaces, particularly \ac{SMI} handlers, which expose relatively standardized invocation mechanisms.
These techniques often rely on symbolic execution \cite{excite}  and domain-specific static analysis \cite{uefuzzer} to recover input formats.
While effective in narrow settings, they incur high overhead, limit scalability, and lack comprehensive sanitizer support.
More importantly, these approaches largely ignore boot-time interfaces, which exhibit richer interaction semantics, complex dependency structures, and diverse input channels, making them difficult to test using existing techniques.
Recent work, such as \fuzzuer{}~\cite{glosner_fuzzuer_2025}, targets boot-time interfaces by executing the full firmware in a simulator and invoking interface functions using synthesized harnesses.
However, as we show in \sect{sec:eval-rq3}, \fuzzuer{} fails to exercise all input channels, particularly those arising from protocol dependencies, resulting in limited coverage and reduced bug-finding effectiveness.

Another line of work~\cite{tanksalkar2025lemix, tay2023greenhouse} explores rehosting firmware as user-space applications and leveraging existing fuzzing tools.
This approach improves scalability and enables the use of user-space sanitizers.
\systemname{} builds on this direction by rehosting \edk{} components using \ac{HBFA}~\cite{hbfa}, while introducing \ac{DCR} to explicitly model dependency interactions as input channels.

\parahead{Harness Generation}
Prior work on automatic harness generation primarily relies on program analysis techniques, such as static analysis \cite{glosner_fuzzuer_2025, uefuzzer}, symbolic execution \cite{baz2015symbolic, STASE}, or their combination~\cite{excite, Yin2023rsfuzzer}.
While effective in generating syntactically correct harnesses, these approaches struggle to capture complex dependency interactions and stateful behaviors inherent in \edk{} protocols.

\noindent\emph{\ac{LLM}-assisted generation.}
Recent works~\cite{yang2025harnessagentscalingautomaticfuzzing, Zhang_2024, PromeFuzz, Liu_OSS_Fuzz_Gen_Automated_Fuzz_2024} demonstrate the effectiveness of \acp{LLM} in generating fuzzing harnesses and improving coverage.
However, these approaches target conventional software systems and do not address the unique challenges of \edk{}, such as protocol dependencies, stateful interactions, and specialized execution environments.

\systemname{} leverages \acp{LLM} for both harness and mock generation, while addressing these challenges through a dependency-aware design and rehosting-based execution.

\section{Conclusion}
We presented \systemname{}, an automated fuzzing framework for \edk{} protocols that rehosts them as user-space applications.
Our key insight is that effective fuzzing of \edk{} protocols requires exercising both interface functions and dependency-driven input channels.
To achieve this, \systemname{} introduces \ac{DCR}, a dependency-aware rehosting technique that combines static analysis and \ac{LLM}-assisted synthesis to automatically generate harnesses and dependency mocks.
Our evaluation on \numprotocols{} real-world \edk{} protocols and \numdrivers{} drivers shows that \systemname{} significantly improves coverage and bug-finding effectiveness over prior work, while discovering multiple previously unknown vulnerabilities.
These results demonstrate the practicality and effectiveness of dependency-aware rehosting for scalable \edk{} firmware fuzzing.

\section*{Data Availability}
\label{sec:openscience}
We will release the following artifacts:
\begin{itemize}[leftmargin=*, noitemsep]
\item \emph{\systemname{}}: the source code, including \metislib{}, our helper libraries, and the prompts used for harness generation, mock generation, and crash triaging.
\item \emph{\ac{LLM}-only baseline}: the pipeline used for the \ac{LLM}-only comparison in \sect{sec:comparison_llm_baseline}.
\item \emph{Multiple \acp{LLM}}: the pipeline used to run \systemname{} with different \acp{LLM}\extonly{ in \apdxref{\sect{sec:eval-rq4-practicality}}}.
\item \emph{Datasets}: all datasets from our evaluation, including the generated harnesses and mocks, fuzzing results, coverage reports, and crash triage results.
\end{itemize}
Our vulnerability commitment ledger~\cite{metisledger} is already public: it records the SHA3-512 hash of each driver's findings on the newly discovered bugs without revealing them, so each finding can be checked against it once it is disclosed.

\bibliographystyle{ACM-Reference-Format}
\bibliography{citations}

\ifextendedreport
\clearpage
\appendix
\input{./sections/appendices.tex}
\fi

\end{document}